# What do astrophysics and the world's oldest profession have in common?[1]

Martín López Corredoira

Instituto de Astrofísica de Canarias

C/.Vía Láctea, s/n

E-38200 La Laguna (Tenerife, Spain)

E-mail: martinlc@iac.es

*Dedicated to Eduardo Simonneau,*
*eye-opener master*

Life is the best teacher, much better than university. I have learnt some things about astrophysics during my last ten years as researcher, but I found other things on the earth that were also worth learning. Trying to understand the mechanics of stars and galaxies is beautiful and I am glad to dedicate my time to this noble occupation. However, one should always bear in mind that we are on the earth, surrounded by other men, and in that respect we must be aware of the floor we stand on and not only look at the sky.

I do not think that astrophysics is a special case within the sciences. In the same manner, I do not think that science is a special case within the world of the administration of the culture. All of us are in the same ship: the world in the capitalist era. Nevertheless, I will concentrate on astrophysics as a human activity since I know this world better from inside (I also know the faculty of philosophy, another "casa de ..." which has more to be criticized about than the institution of astrophysics, but not here). In the present paper I want to tell of my impressions about this world, which I know from close observation, in an open way and without self-censorship. I want to tell the facts as I think they are, without worrying about whether this is nice for the reader or not, or whether this can be published somewhere. I think the only method of reaching the truth is to never be afraid to tell the truth, and to put it before other interests, such as prospering from the system, getting a position, publishing in prestigious journals, etc. Probably, there might be errors in what is said here, since they are my subjective appreciations. However, they do not matter when the goal is important: saying honestly what one thinks.

---

The way the actual institutions work is very complex. One begins to understand them once one has worked some time in them; inside them. I think that a philosopher or a sociologist cannot correctly infer a plausible theory on social mechanics in the sciences by reading four books, without any direct knowledge of what is being 'cooked up' in these institutions. Perhaps, it can be inferred by comparison with other institutions, but not from simply reading some books, since there are practically no publications that reflect the real situation. Truth is not always achievable through a bookworm, since some truths are not written or their diffusion is very limited (as may be the case with this text), or perhaps because somebody has an interest in them remaining unknown.

**Students**

The first contact with research takes place when one prepares a PhD thesis. Here, as in many other disciplines, the system adopts a clear position: "things are as we say; either you take it or you leave it". If one wants to work on the research, one must be in the service of a program that is predetermined by the authorities responsible for the system. If the student wants to get economic as well as department support, then his or her role must be obedient to, and assimilating of the traditions of the department.

In a jokingly ironic and cynical sense, and with a certain presumption not so very far from truth, it is usually referred to as student "slaves". The slaves are in charge of doing the most monotonous research tasks (observing during long periods through the telescopes, data reducing, etc.) in the service of the team where they work. There also exist in the hierarchy of the system some inferior figures: the temporal (for a few months) scholarship holders; they are normally pupils who have not finished their career and, therefore, they are under PhD students. Those are usually called 'summer slaves', because their contracts are in force during summer months, and there is not enough time in such a period for them to learn something about research. Therefore, they are used as cheap manpower: a few days to learn a mechanical task and the rest of the summer to apply its routine.

This appraisal of the treatment toward students is really not always applicable. In my case, for instance, it was not. However, I feel certain that the reality of this exploitation reality is quite extensive and rather more common than would be desirable. Of course, I am telling what several researchers have told me rather than describing statistical data published by some official organization. Nevertheless, I judge that the sources are sufficiently representative.

In some cases, those students that do most of the work cannot write their results in a paper, but the bosses do it instead, as first authors of the paper. Students are told that they do not know how to write their own work. In other cases, when the supervisor sees that things do not have the outcome he or she wants, the supervisor abandons the student. In some cases, the supervisor steals the ideas of the student. In other cases, the granting of the PhD to a student is over before the thesis is finished because he or she was exploited by having to perform other activities aside from the writing of the thesis, or the boss had no time to attend to the explanations produced by the student. In such cases, the student must struggle to survive while finishing the work.

Few bosses sit down and work with students. Normally, they spend some time during the early days to explain how to do things. After that, the student must do the routine tasks. The boss just gives the ideas, if they have them; otherwise, just makes minor corrections. The student spends weeks or months in front of the computer, fighting with a program that does not work, with annoying calculations or simulations that consume a lot of time. Students spend complete nights at the telescope (the boss is usually present as well, but only the first time to explain to the pupil how the machine works, or when there is a novelty or extraordinary observations unrelated with the usual routine. It is very normal that they go to sleep at midnight and leave students with routine work till the dawn if there is nothing holding them there), for complete weeks at the telescope. After that, he goes down with the tapes of several gigabytes of data, and reduces them; that is, processes them, and extracts information from the observations. This task usually requires several months. If there is some complication due to an error in the procedure, and the reduction must be repeated, it will take longer.

Meanwhile the boss manages and puts ideas. The boss will say: "All right, but you could do what's his name, or that, or this thing beyond that". The student will spend one week to do what's his name. He will spend two weeks to do that and finally he will realize that it is not feasible. This thing beyond that is surely a stupidity but the boss cannot be convinced of that until it is checked with some calculations (of course, carried out by the student). Finally, he delivers the results to the boss, and he says: these are nice but I prefer it as they were before.

Which objection addresses this situation? Is it not normal by any chance that the master teaches the student, and the students effect what they are ordered to do while learning? Certainly, it must be so. The fact is that the recently graduated student does not know much about the specific area of work, and must be brought up to date. Nevertheless, they are not novices without knowledge. Usually, they have more general knowledge about astrophysics as a whole than the specialist who knows much but only about their own specific area. Moreover, students have some advantages over the master in this case: they are more creative, more open—with less prejudice—and can give new and fresh points of view about research to develop, instead of following anachronistic traditions that are embedded in the interests of the person who has spent a whole life with one idea. PhD students can produce ideas if allowed to produce them, even away from the track that was predetermined for them. In this regard, it would stand more to reason that the monotonous work should be in the hands of those with exhausted creativity, those aged, reputed experts who will produce nothing but copies of what they have always produced. However, the world of science does not stand to reason but that of power: the captain gives the orders to the sailor. Because of this, it is usual for routine work to be carried out by students. This is not so that they may learn (since one learns the first time, but not by doing the same task a hundred times), but in order that they produce. In some few cases, PhD students do their own research along with their duties with the supervisor, and in fields other than the subject of their thesis (I would recommend this to the future students), but that is not the most common way.

I was once told an anecdote with regard to this. I ignore whether it was true or not, but it seems that it is a real case: a student talks with his supervisor and says "I had an idea". Then, the supervisor replies: "Ah! You have time to think?". Is this the way to form future scientists? Having them spend time in a thousand routine tasks without free time to think freely? Thinking how to corroborate, yet again, with an idea that originated from a specialist of repute is to be encouraged. However, spending time to think about one's own ideas, without permission, is something that is really not encouraged by the system, quite the opposite. Initiative is discouraged with arguments such as what is established is well established. Workers for science instead of thinkers are created.

What is the objection in this situation?—I continue to ask. Mainly, that creativity is not taught but industrial (mass-produced) science, and the period of optimum creativity of a scientist is exhausted with these ups and downs. We must take into account that, in the long history of science, the majority of great ideas were produced by young scientists. If young students, who could potentially produce new ideas, are used as slaves (or perhaps it is better to say 'workers for science'), then perpetuation of the ancient things and stagnation of intellect is rife. Even though, apparently, if one were to believe the mass media, a scientific revolution is being produced every day.

**Postdocs, permanent positions**

The researcher who wants to live on his work in the research world must aspire to a permanent position, that is to say, to become part of the body of functionaries, well known by all because of their efficiency at all levels. Beside jokes, the fact is that the life and motivations of the postdoctoral researcher are marked and oriented towards the obtaining of the position as the ultimate purpose. In the best of the cases, the functionary will still be motivated after the goal is obtained, but in many other cases the opposite thing happens.

A student who has just read the thesis obtains rarely an immediate permanent position, but he/she must before go through several institutions (some of them necessarily in a foreign country, although there are exceptions) with temporal contracts known as "postdocs". I think really that this is one of the cleverest things the system has, because at least the researcher has some years of his/her life to look for their own pathways in research and, at the same time, avoid the early stagnation that is usually produced by the early permanent positions.

"Postdoc" status constitutes a hierarchical position of workers for science above the PhD student, but below the researcher in a permanent position. A large part of the routine scientific tasks falls to these figures but to a lesser extent than the PhD students since, in quite a lot of cases, they have their own mobility. In many other cases, they are contracted workers for a predetermined program. I think that travelling to a foreign country or not is of little relevance. Perhaps it is important only to become more fluent with English or other languages. Nowadays research is quite global in nature. Few things can be learnt in a country that cannot be learnt in one's own. In all places, the same thing is told and the same thing is done, with minor differences. The type of research to take on a contact has, perhaps, a larger influence than does travelling to a foreign country. Generally, after completion of a thesis, the researcher continues to be surrounded in the same type of environment, so he/she does not learn much that is new.

It must be said that not all doctors continue their careers as researchers. In many cases, motivations from private life are obstructions in the requirement for mobility in the job. Also the high competition necessary makes it possible for only a few to continue. In order to gain a postdoc position, a good curriculum is necessary, which is not necessarily associated with genius but the capacity to work, and the support or recommendation of somebody within the system. Without the appropriate recommendation, a career may be truncated. Hence, it is necessary to look for congeniality in the world. A way to look for congeniality is by following the stream of general trends in research without trying to create a critical front. With regard to the curriculum, the major weight normally comes from the number of publications in professional journals. I say number (quantity) rather than quality because the predominant parameter is really the first one. Quality is valued when the committee evaluating the person who is applying for the position is a specialist in the same field and with the same ideas (that is, not a competence in defence of other theories). Since each specialist thinks that his/her field is the most relevant, the curriculum will always find support with respect to quality when it is oriented towards the interests of the judging tribunal. Otherwise, it will be just a number, evaluated with weight, as with school tasks.

This way of evaluating the efforts of a researcher will be a constant throughout the researcher's life; either to secure a position, or to obtain telescope time, or to fund money for a project, etc. Later, I will talk further about these questions. With regard to the postdoc positions, we must see in this evaluation system an indirect pressure over the putative free choice in research in order to focus towards those already given. The number of publications with little critical content, as well as short-term congeniality, is an indication of this pressure. These will be the factors used to secure other postdocs, or a permanent position, if the person does not leave the research before, or chooses to earn his/her living in another way.

In my experience, for instance, although not many crashes have taken place, I have being working in some non-orthodox fields and, consequently, I realized the problems that arise when one works in a field that was not recommended. Due to my approach to some researchers who work in areas of scientific theory with little orthodoxy in order to discuss certain data or exchange opinions, I had to listen to much advice. They claimed to have gone away from these research fields and all possible relationship with the aforementioned researchers. The rest of the community could relate to their position and this would be an obstacle in securing a postdoc or permanent position in the future.

When I was invited to give a talk about the topic in a certain institute, a senior scientist said to me that giving such a talk would mean forgetting about obtaining any position there. This is not blackmail, but it is close.

From my own experiences and those of others, I have deduced that doors are opened and offers made to those who are servile and uncritical. A lot must be produced, but without great aspirations to say something important. It is sad to have to say that the positions I obtained were given thanks to the works I consider less relevant, while those works that I consider interesting have created problems for me, together with much discussions, headaches, and inattention. It is sad, but it is so. I am certain that I am not an isolated case.

**Publications, referees**

The fruits of scientific activities are gathered in specialized journals, those journals that will be read by other specialists and distributed in libraries of the research institutes all over the world, with exorbitant prices, either to publish or to receive the journal, which can only be afforded by the wealthy institutions. Of course, it should be mentioned here that the scientific journal business is not a trifling question. Nowadays, the journals are a powerful communication tool, written in the international language of English, and with enviable accessibility. It must be recognized that the present-day system of scientific publications is very superior to that of other cultural fields, where the unification of the language and the publications is spread over many local journals with difficult access.

As it is well known, control of communications and practice of power are closely related. I do not think that I have discovered anything new with such an affirmation. Thus the system, far from allowing free publication of results among professionals, works hand in hand with censorship. Theoretically, it was conceived as a quality control but its functions are frequently extended to the control of power. Those researchers who want to publish in these journals are subject to the dictates of the chosen referee and the journal editors, who will say whether the paper is accepted or not. The referee, by choice, is usually anonymous. I had even a case in which the editor was anonymous as well, and we only knew the name of the secretary. This fact points out that the activity is not always honest. If it were honest, the referee would not hide himself behind anonymity or, perhaps, be afraid to be pointed out as the disparager. If somebody thinks they are giving good advice then nomination of their work for a journal should not be hidden.

Generally, papers are submitted to referees who are experts in the matter. They can afford their knowledge to improve the quality of the paper to publish, or detect errors in a calculation, if any, or detect contradictions with some data, etc. In principle, the idea is good, and it would be better if the refereeing process were always objective and impartial. I think that it is not the case. There are many cases in which the fate of a paper is dictated by a conflict of interests rather than the merits of the paper.

From my experience in the publication of scientific papers in refereed international journals, I have observed that the reports of the referees rarely detect errors in calculations or data reduction procedures, because the referees are not patient enough to carry out the calculations again or check the codes. Apart from minor details—changing a plot in order to see it better, explain a paragraph better, cite some other paper (in many cases the referee advises to cite some paper of their own, or by collaborators etc.)—, objections very often are to do with the referee's own opinion or how convinced they may be about the contents of what is going to be published. Generally, according to my experience and other experiences that I could list, the more controversial the topic, and the more challenging it is to established ideas and the newer the approach, then the more difficult the problems will be in publishing it, and the higher the probability of being rejected. Otherwise, when one writes a paper that repeats what has already been said by hundreds of papers on the same topic—with some changes, perhaps, in the parameters if a theoretical model, or focusing on different objects than those which have already been observed or observing the same objects with best data—and reaching the same conclusions which are already known and in agreement with everybody (especially the referee, who is usually representative of orthodox ideas; an exception might be in relation to some secondary points), in these cases, a referee will be more likely to be less belligerent and may even send congratulations to the authors.

The background problem is as follows: referees are persons who have dedicated their whole life to do research in the few problems of a particular field. They are widely recognized persons in their field and their social status is due to their contributions in the field. As persons with experience and prestige, and sometimes associated with an excess of vanity, they usually think along the lines of 'I am a great specialist in this field. I know the interesting and crucial ideas about it. If a new idea were presented, either it is of little interest, or it is wrong, or I would have thought of it before. Therefore, if somebody presents a new work that tries to tap into crucial questions, either it is a continuation of my own work and ideas and those in which I was involved, or it is wrong'. Moreover, it might be misconstrued as competitive (argumentative, contrarian) for somebody to publish a theory or interpretations different to those argued by the referee. Perhaps it is somewhat exaggerated to attribute this thought to many "authorities of a field". However, I think that something like this thought is more or less present. Of course, this vanity is not explicitly recognized. The fact is that this psychological mechanism, although not explicit, can be present in most of the cases in which there is a discussion about the credibility or how convincing is a theory, or any other subjective approach. Certainly, science has an objective content, and the data and maths are there, independent of what is thought about them. However, the data interpretation and the plausibility of theories is something that is subject to the human factor—beliefs and, in many cases, prejudices. This carries much weight in the censorship of scientific publications. Of course, I must also say that there are many good referees who do a wonderful work as well.

What is the consequence of this? This really has a positive effect: avoiding the publication of hundreds or thousands of incorrect papers with absurd ideas that have no sense. Nevertheless, the negative side is also obvious: obstruction of thought and the few interesting ideas that could be produced. Well-done works but without ideas, works of specialist artisans, are rewarded. Creativity is damned. It seems that the system gives the message that no ideas are needed. It seems the system, with the set of higher authorities, is saying that astrophysics or any other science only needs to work out some details. It is accepted that the basis of what is known is correct, the present-day theories are more or less correct and only manpower is needed to fit some parameters or aspects of minor importance. A Copernican revolution is totally unthinkable within the actual system, even if truth were different to present-day theories. With regard to this, there are not many differences between the present-day academy and the university in the sixteenth and seventeenth centuries that conformed to the Church and Aristotle's texts. It is not true that science has similarities to an ancient religion, as has been charged on many occasions. Scientific arguments are very different to religious arguments. Nevertheless, the behaviour of human groups that claim to have the truth among their hands is very similar.

An important exception to this censorship is the existence of the electronic preprints "astro-ph" (or other names for other fields in physics or maths). This is, in my opinion, the most important effort to open doors to the research in which the originating author can place a paper without the control of a referee[2]. Normally, papers are put in astro-ph once accepted in a major refereed journal, but the author can also put the paper in astro-ph before it is accepted by a journal. There are also some printed journals without censorship, but these are minor journals that are practically unread. Only astro-ph has a large diffusion. The counterpart of this freedom of publication in astro-ph preprints is that the papers are not officially recognized until they are accepted by a major journal. They cannot be used to support any proposal of time application in telescopes, for instance. They cannot be considered as papers to argue an idea against other approaches. They can even be ignored as if they did not exist. If a leading specialist is asked about a paper in astro-ph that is not accepted in a major journal, the specialist can simply reply that the paper was not published and, therefore, can ignore its contents as if it did not exist (this happens with many accepted papers too). Nevertheless, in my opinion, the astro-ph preprints are a good tool in the research. At least, another researcher without prejudices can read the paper and judge its quality. It is also useful for the author: in my case, I have found important errors in some papers thanks to the comments of a person who has read it in astro-ph; errors which could not be found by any referee, who are used to read a paper superficially, and reject it with a simple rebuff without good arguments when he/she does not like it. Of course, there is a lot of rubbish in astro-ph but there is a lot of rubbish among accepted papers in leading journals too. Somebody could also steal ideas, but that also happens with accepted papers that went unnoticed at their time and years later an author of prestige rediscovers them, and takes on the ideas. How many authors of the old Soviet Union have discovered many interesting things, which the world could not know until a clever North American researcher, with plenty of dollars, rediscovered it and Queen Ann's dead![3]

---

[2] Update: Regrettably, in the beginning of 2004, two months after I posted this paper in arXiv.org in the section of astro-ph, a new policy was established which curtails the freedom of dissemination of ideas. The new system requires that, if somebody who has never posted a paper in a section of the archives arXiv.org wants to do it, he/she must get the permission from a scientist who uses it regularly. Even if somebody uses often a section like astro-ph or quant-ph, he/she is not free to post papers in other sections unless he/she gets a permission of a peer after the revision of the paper to be posted. In the last years, the censorship system is becoming more refined: it is removing as possible peer reviewers the names of the people who give consent to post papers which are not welcome by the establishment. In my case, after giving support to some person who wanted to publish some challenging ideas (I thought these ideas were wrong, but they were worth to be published), I was informed that I cannot be a peer reviewer anymore.

Another problem is the number of papers. In the astrophysics branch, 30 thousand papers per year are published. This is a very high number, the reading of which cannot be undertaken by even the most hardworking of readers. Within a restricted sub-field of astrophysics, such as comets, Seyfert galaxies or others, one can find 500 or 1000 papers per year in relation to the topic, which is still a huge amount even if only to have a quick look. This number has grown up and continues to grow in an uncontrolled way over time. Chandrasekhar, one of the old editors of "Astrophysical Journal", after leaving his duties as editor, realized the increasing overflow in the number of papers per year. He used to say, ironically, that the increasing velocity of the paper number is higher than the speed of light, but there is nothing to worry about for there is no violation of any physical law because these papers carry no information.

Since most of these papers do not contribute anything important to the field but dispensable details, the possibilities of the few important papers that undergo censorship that would otherwise have an impact on the community are significantly reduced. This means that, once the obstacle of direct censorship in the journals is removed, the researcher who hazards new ideas will have to fight with an indirect censorship: the superproduction of papers that hide what is uninteresting to the system. Propaganda is the key element for a paper to become known. For this, the leading specialists again have the advantage because they control most of the threads which move the publicity machinery; they have the appropriate contacts, they write reviews (summaries of scientific discoveries within a field), they organize congresses and give talks as invited speakers. Moreover, the reproduction of standard ideas has itself much more acceptance because the interests of those who work with them are many while the diffusion of new ideas is interesting only to their creators.

---

[3] Update: I can give a recent example with the research I developed in Tenerife (Spain) within the TMGS team. Between 1994 and 2003, our group has been publishing some papers on the existence of a long bar in our Galaxy with some peculiar characteristics [see, for instance, Hammersley et al. (2000, Mon. Not. R. Astron. Soc. 317, L45) or López-Corredoira et al. (2001, Astron. Astrophys. 373, 139)]. In 2005, a group of U.S. astronomers associated with the mining of the "Spitzer" satellite data published a paper about the discovery of the same bar with the same characteristics: Benjamin et al. (2005, Astrophys. J. 630, L149). Moreover, they produced a Press release with the title "Galactic survey reveals a new look for the Milky Way" which was divulged in many mass-media outlets. A "new" look? No!... this was proposed years ago by us. Benjamin et al. do not cite our works when they talk about the bar. According to some information which has reached us, they cited us in a first version of their paper (so they knew us; it is not a question of lack of information) but they decided to remove this citation and talk about their discovery as it were something new in order to save space.

**Congresses**

The phenomenon of congresses, symposia, workshops, schools, meetings or any opportunity of joining other professionals to communicate and exchange ideas, has become widespread. The phenomenon is not only present in sciences but in any professional environment and has increased hugely during recent years. In the first decades of the twentieth century, while every month discoveries of huge importance were being made for the development of physics (for instance, in relativity, or quantum physics), such gatherings were celebrated once in a blue moon, with important international congresses held annually, or at longer intervals. Nowadays, in astrophysics alone—one of the multiple branches among all researches in the physical sciences—around 200 or 300 international congresses are celebrated per year, apart from small local or national meetings. The saddest aspect of the question is that the conceptual level of development of physics today is far below what was reached in the beginning of the twentieth century.

Holidays can be a reason to attend congresses. Many of them are celebrated in exotic or tourist destinations, which allow leading scientists and their friends to enjoy a holiday with public funds. However, the main purpose of congresses is not to promote tourism but the diffusion of the information in a micro-field of astrophysics (or any other science), and trying to give a wider, more global overview to a given topic. In order to do that, the congress is usually structured into long series of talks that last several days. The invited speakers are highlighted and they are allowed to give long talks, of up to one hour, to talk about their own research or those papers in which they are interested. They comprise ten or twenty leading specialists who are friends of the congress organizers or share kindred ideas. There are also selected speakers who are among those who apply to give a talk. Since their number is high (around fifty in a congress of three or four days without parallel sessions), the spare time is distributed, so that each of them can speak fifteen or twenty minutes. In this short time, they must discuss their research activities during the last two or three years. Consequently, the result is "concentrated" talk sessions that quickly exhausts the attention of the audience. Basically, they have the utility of propaganda. It is useful to say that I have carried out a work about this and the one who wants to know something of it must read my paper. Finally, there is a room for posters, in which hundreds of condensed papyruses concentrate texts and figures in a square meter of bristol board per poster to show results and obtain propaganda value from them. This reminds me of the trade exhibitions in which each company shows their merchandise for publicity. Moreover, publicity resources to attract the attention of the assistants at the congress (the same persons who show posters or offer talks) with pictures and poster designs of showy colors, videos of numerous simulations or films of some impact in talks (there are even cases of researchers who pay to professional animation creators such as "DreamWorks" to make the videos), etc. All this has the goal to attract the attention of the audience who get lost among the tons of information; dispensable information since there is not much new to tell at each congress, simple technical details without too much relevance. The battle of the scientist is not finding new good ideas, but finding the way to sell mean, unworthy ideas. Marketing is more important than Math. It is all just publicity, and meeting colleagues to talk about, and discuss future collaborations.

This publicity is very important for the system and for the control of information flux. And it is important to give priorities in the congresses, provided that the first purge was already carried out in the research institute of the scientist. The scientist must first convince their own institution that the expense of assistance at the congress is justified by the contents that will be shown in order to secure a subvention. Sometimes, the assistance of certain personages is forbidden. For instance, in the conferences about cosmology at the Vatican in 1982, only the staunchest defenders of the Big Bang theory were allowed to participate, and marked individuals were left aside long with the defenders of opposing views such as Hoyle, Ambartsumian or Geoffrey Burbidge. As told by the physicist W. Kundt, *"[cases] where I was not invited to topical meetings, and even where I was sent home from a meeting on the day of my arrival"*. Fortunately, neither the Vatican example nor the experiences of Kundt are too extensive. There is a filter, but, even so, the censorship system is less efficient than the refereeing of journals and audacious theories pass censorship more frequently than is the case with the journals. Of course, when the proceedings book (which contains the written texts of talks and posters) is published, some authors can use only two pages in small format, or none, while others take up thirty or forty pages.

**Financing, astropoliticians and supervedettes**

Among senior researchers, not all of them have the same weight or authority in the hierarchy. There exist, as in any University department, certain ranks such as position in a chair, department director, etc. Apart from these nominal ranks, there also exist certain power status indicators that are associated with other factors.

Many senior scientists and functionaries with security of tenure devote most of their time to teaching at universities. Perhaps they take a 'slave'—I mean a PhD student—to carry out work which the senior scientist will then co-jointly sign, in order to show that he actually does do some research. In first rank institutes, competition is higher. In those institutes, some of the researchers are leaders of a project, and pursue 'impact' a particular one, that is, the project have the objective of producing many published papers and they command a certain respect from the specialists.

The project's main researcher is the leader of a group with several PhD students, several postdocs and, perhaps, some senior scientist of lower status. There are even cases in which this main researcher may have all the postdocs of a small institute. This main researcher is usually a type of commercial manager, where certain elements could be termed agent and adviser. I call them "astropoliticians" and have known several of them.

I think that most of astropoliticians exhibit a similar behavioural pattern. One must make an appointment to simply talk with them, since they are always busy with a thousand and one tasks. "I have no time" is one of the favourite sentences of the astropolitician, a man of our era. We live in a time in which even the pipsqueaks pretend to conduct themselves and take over as if they were important men (a minister or someone who is very important) and deliver the self-important response of "I have no time" or "I am busy"'.

Within their offices, it is usual for them to receive three or four phone calls in less than thirty minutes. They receive tens or hundreds of e-mails daily. When an appointment is required in order to present some scientific results for an opinion, the astropolitician has to revise their agenda, mentally or in a notebook, because there is always a meeting to be attended somewhere. In addition, much travel is undertaken both nationally and to foreign countries. They must prepare talks, because they are the main speakers at the various congresses. They must attend a large number of meetings of astropoliticians to obtain agreements (scientific collaborations, not commercial agreements, but the outcomes are similar), or negotiate some budgetary entry, or create propaganda for the project in order to obtain some economic benefit or achieve an impact in some other way, or to think up—together with other astropoliticians—yet another macro-project that will cost many millions of euros and will employ the many researchers in yet more monotonous work. Of course, they are not these researchers but the persons who are subjected to their orders—along with other new slaves who will be brought on stream with the money received from various negotiations. When astropoliticians are not travelling or in a meeting, they usually are busy with the preparation of periodic information bulletins concerning project activities or filling in forms to apply for new telescope time (which may also be done by students) or applying pressure for economic support for the project (travel, computers, scientific instruments, etc.) on some ministry or other for various types of assistance. In their spare time, they usually are busy with the coordination of project staff and their work efforts as well as establishing research priorities. The hen takes a cup of coffee to rest from the bureaucracy duties, while the little chickens are all around, eager to show their results. The astropoliticians listen to (in many cases, they do not listen to), and read papers by low-ranking 'workers' and express their opinion and, more than likely, suggest changes according to their prejudices. It is a rare day when the astropolitician may sit down to do some actual scientific work in the true sense of the word "scientific". Perhaps, they may dedicate a few hours on some days to teach an aspect or some feature to a low-ranking science worker. In most cases, however, it is not they who dedicate months to work on resolving various problems but, instead, their PhD students or postdocs.

When an astropolitician is able to display particularly bright attributes as agent and trader for the science that created these workers, we have an example of a "star". A star that is dazzling in both its brilliance and prominence. In other words, a "supervedette", the great star among the stars. In an institute with more than a hundred researchers, there are usually only one or two supervedettes. Their identification is not very difficult because they are an essential reference of the particular institute, especially in the image given to the world outside. If a journalist visits in order to write an article about what is happening at the institute, the supervedette comes to the fore. If his/her team does any work, the press is quickly called in to announce to the world what So-and-so "et al." (that is, "and collaborators", although the name of the low-rank workers who has made the discovery is usually not important), with the phenomenon of fame being confined to the supervedette. They have usually a good eye for choosing the topic that has high popular impact (not necessarily topics of high scientific importance). If the topic is without fuss, they will announce it with a lot of ballyhoo in order to create a fuss. They publish without difficulty in the journals; they write professional and popular books. They are the owners of the congresses together with others of similar ilk, they get all the telescope time they want, and the budget for their activities is gargantuan. These persons do not think in terms of minor but major goals. They lead large multimillion-euro projects. In addition, a single phone call can translate into widely disseminated propaganda via journals and television. They are in national and international price competition as the 'best researcher'. They are mixed with famous high society personages. Circumstances may vary depending on how big is the "star" but he/she is—in short—a basis of envy for any astropolitician.

Of course, there are exceptions to this behaviour. Any attempt at generalizing a behaviour pattern of a given collective is always subject to the corresponding corrections for the particular details of each case. There are some cases in which a senior scientist, even a leader of a project, works at the same tasks as do lower-rank workers, and does not dedicate too much time to administrative tasks. This, however, is not the most customary scenario.

I have the particular case in mind of a senior scientist with secure tenure that is a good example of such exceptions. This person does not lead any project. He works with his own ideas, or together with a collaborator. This person goes to very few congresses (perhaps one every five years or less). He does not go to meetings. While talking with him in his office, he has few phone calls. He spends a short time in answering e-mails. He always has time to receive visitors to his office when somebody wants to talk with him. Apart from working as a good professional in his field, he has a wide knowledge of many subjects and many other fields of study. This makes it possible for people to speak with him about a physics topic or on any one of a number of other fields, since he knows a great deal about philosophy and history and has a very good memory of what he reads. He is accustomed to thinking. In fact, on many occasions when visiting him in his office, I found him actually thinking; not performing some task related to administration or self-promotion and talking on the phone, but thinking. When we speak, I usually discover, in the lucidity of his thoughts and reflections, the answers to many aspects of astrophysics. He thinks quickly (perhaps too quickly for a listener to understand what he is thinking as his thoughts are being articulated) and is almost always correct. He has great intuition and visualizes a problem in order to discern its elements. His help is always likely to be given when one has difficulty in solving a problem in physics. This person is a prototype of the learned scientist and is quite rare these days in our professional scientific jungle. He certainly is not mainstream and, in fact, is considered a second rank scientist. Few people know of him outside the institute, and his complex works are nearly always forgotten because of the absence of any accompanying propaganda. Nowadays (and most likely in the past as well), the 'trumpeter' is another species of scientist: the executive with attaché case in hand; the professional science agent. In many cases, this agent does not know how to think about solving a scientific problem, nor is there too much insight or knowledge about physics and astronomy. Many findings by members of astropolitics do not deserve to be highlighted from an objective standpoint, but this objectivity is difficult to achieve with all researchers believing their own works are important. Therefore, the astropolitician is the triumpher.

**Press, television, propaganda**

As mentioned earlier, press, radio, television or similar media, are useful tools for the manipulation of information and mercantilist propaganda. The knowledge of society in relation to scientific activities in general stems almost totally from press, television and propaganda sources. Therefore, control of these media is an ideal mean of achieving what the controllers of a society desired to have perceived or believed by the general public.

Most journalists responsible for writing articles about science have little knowledge about what they write; perhaps they have some knowledge on science in general, but they are very far from being in control of all the existing specialties. This is the situation in even the most prestigious newspapers in the country. For the less prestigious ones, it is even more likely their journalists have no scientific culture whatsoever. Because of this, the journalist is obliged to believe what the researcher says. If they are told that a great impact discovery has just been made, the journalist must trust that it is so, since the journalist has no personal knowledge from which to cast doubt on the veracity of the researcher's statements. The determining factor in these situations is the researcher's reputation. Thus, fame feeds fame: a prestigious researcher is usually surrounded by a swarm of journalists. The propaganda they distribute will contribute to increase the 'fame' of the researcher. In this regard, there are not many differences between the 'fame' achieved by a scientist and that attracted by a singer or a protagonist of the pink press: it is all question of availability to the mass media.

Researchers perhaps overestimate the value of their own work, but do not usually deform or exaggerate, or say the opposite of what it is—at least not intentionally. The journalist does these things and does them intentionally. The goal is the impact, which is something of high value among friends of misinformation and ballyhoo. It is, apparently, what they are taught in the faculties of journalism. Thus a good deal of the information published in the press about recent scientific discoveries contains significant errors and receives appraisals which are totally inconsistent with the purported newsworthiness of the reported item. Titles--headlines often distort the news. I still remember reading in a newspaper something like "extraterrestrial mummy", in reference to the fact that some researchers had found the tomb of an ancient Egyptian pharaoh that had been built with stones from a place where an extraterrestrial meteorite deposited on Earth in the past. Incidentally, it seems that those who are ignorant of science are usually worried about extraterrestrial life and, therefore, journalists feel their duty is to satisfy the readers who are eager for news related to the subject. Perhaps because of this, there is much news published about the discovery of new extra-solar planets. This usually presents the news item as if it were the first time an extra-solar planet had been discovered, instead of the true facts that are that tens of them have already been discovered and all with masses thousands of times greater than the mass of our planet. These 'masses' have nothing to do with earth-like planets. And, no! extraterrestrial life has not, as yet been discovered—the most common question that journalists make, when talking with the populist mouth, in their eagerness to convert science into a side-show.

The number of cases where scientific news is published with a disproportionately huge amount of ballyhoo—such as Einstein's theory of relativity is no longer correct, or that there is life on Mars, or cold fusion, or similar is very high. In many cases it stems from the interpretations made by journalists because they do not understand the subject. In other cases, the sources may be real discoveries that have been published in scientific journals, but which are still being discussed and about which certain controversies remain. After some months of the sensationalistic publication, the scientific community usually clarifies that the discovery was not such a discovery because there were some errors in their results. However, general public only remember the huge ballyhoo, not the reply that refutes it. Apparently, the truth is not so interesting for commercial purposes, and does not help to sell further newspapers or magazines. If somebody wants to know something about science, I would advise not to do it through the press or television, but through textbooks. I would also advise them to forget the newspapers. Future will tell us what is being done right now.

In spite of the imperfections of scientific communication throughout the mass media, it remains the fundamental pillar of the relationship between scientists and society. Many of the subventions of multimillion-euro bequests depend on it. For example, the case of the Antarctic stone with a life of Martian origin was famous all around the world. It gave rise to a large subvention for further research into the topic from the American government. Afterwards, the news was denied—the stone was contaminated with terrestrial life—but those who got the money for the project had already obtained what they wanted. Incidentally, the paper about this discovery was published by 'Nature', a professional journal of prestige. The paper itself had been submitted to three or four referees and accepted.

In some cases, the opposite thing happens. Subventions are not a consequence of the press, rather the press is a consequence of subventions. When large amounts of money are invested in a project (the sum may be as high as hundreds of millions or euros or even billions of euros), justifying the investment of public funds becomes necessary. Therefore, the press is usually called in to explain to the nation the great discoveries obtained, with thanks to the taxes paid by the nation's taxes. This, again, is propaganda. In some cases there are somewhat more important discoveries, but in many cases there is nothing interesting. Specifically, in the last examples the press is needed to exaggerate a matter and create the belief that the items newsworthiness is more important than it actually was. Things are said, such as so and so many new galaxies in the Universe have been discovered, as if the old surveys had not encompassed millions of galaxies yet. When an ignorant general public, most of whose members do not even know what constitutes a galaxy, reads such news, they become convinced of the greatness of the stated venture undertaken by the survey.

Moreover, the press is not always at the service of all-important scientific phenomena. Without fame, without money and without the recommendation of, or support from, a prestigious team of researchers, even the best of scientists, working in the most important fields, would be not listened to, nor paid any attention. Thus, yet another factor arises to account for the isolation of the non-mainstream scientist.

"An individual with few resources getting what we could not get with billions of euros. This would be a scandal, and we cannot allow it". This is the message of the actual capitalist society where money imposes its power.

**Telescope time**

Astrophysics, as with all sciences, has a theoretical part and an experimental/observational part. At present, it is restricted to the observation of nature. In this science one can see but one cannot touch. For obvious reasons, experiments cannot be conducted with astronomical objects. There may be great advances in research due to purely theoretical work. However, it all depends finally on the contrast between such theoretical work and observations. In order to be successful, a theory must be able to predict certain phenomena that other theories cannot explain. Even Einstein's general relativity had to await the observational confirmation of the starlight being bent by the gravitational field of the Sun in order to have the impact it had—it was measured in 1919, by Eddington et al. Indeed, Crommelin, one of the lower-ranking of co-workers, together with other collaborators from Brazil, carried out the higher precision measurements, while Eddington's group in the Spanish Guinea had severe weather and were unable to obtain such precision in their measurement data. Therefore, advances in astrophysics advances are closely related to observational advances.

At the beginning of the twentieth century, astrophysics, and science in general, had made important progress due mainly to the search for new ideas by several famous researchers. When Hubble and Eddington were asked what they expected to find with the new five-meter telescopes that were going to be built, their reply was "if we knew the answer, there would be no purpose in building it". Nowadays, however, the situation is very different. Before using large and even not so large telescopes, a tribunal of specialists must be convinced that something will be found which is already expected. In order to use these great installations, some forms must be filled up between six and twelve months before the observation date. In these forms, one must clarify what finding is anticipated as a result of the application of their measurements and observations. The tribunal of specialists must be shown the purpose of pursuing the observations. In addition, profile data on the researchers must be filled. Of course, the greater the history of observational publications and telescope time that a researcher has had in the past, the higher the probability of gaining further telescope time. Therefore, the researcher will be able to publish more papers than anybody else, although all the papers are similar and without any worthwhile, or new ideas – but prestige will be increased and enhanced. Using a large telescope or satellite to obtain data also adds prestige to the published results. One says, for instance, "data obtained with Hubble space telescope", and this serves to presume that these data are far worthier for science than any other information gathered with less prestigious telescopes. It is a circular loop, and one just needs to establish a certain level of congeniality with the established leading scientists in order to enter that circle. From my own experience, and another whom I know, in the sending of proposals the probability of gaining telescope time increases very significantly when one of the co-authors of the proposal is among the members of the tribunal (although that member will not actually be judging the particular proposal).

Nonetheless, the biggest problem for the advance of science is that new ideas are not welcome among the tribunals that authorize telescope time. If somebody applies for telescope time in order to test the predictions of an alternative theory, rather than the standard one, the proposal is most likely to be denied. We are not talking about amateurs for whom some crazy idea has occasionally gone to their heads; we are talking about great professionals whose only defect is in doubting the ideas which all the rest of scientific thinking considers untouchable. The system does not support an ideological plurality within the science. It is said that there is freedom in research, but this is just a lie as are so many other statements made by politicians, ostensibly in the name of the democracy. Of course, anyone can think what he wants to think, but the installations, the prestigious publications and the propaganda are only for those who want to make a science a reconfirmation and underscoring of certain prejudices, rather than an opening towards new horizons.

It might also happen that somebody presents an idea or an objective that is interesting for its study, but the work cannot be developed because the tribunal does not make telescope time available.

Immediately, people from the tribunal with further resources, seeing something interesting in the idea, begin to develop it and make the discovery their own.

All this is understandable, although not acceptable—at least from my standpoint. The scientific body politics is convenient for large flows of capital. A telescope such as the 10-meter one that is being built in La Palma with a Spanish budget costs the huge amount of around one hundred million euros. If one looks for high amounts of money, space telescopes and the great satellite research projects costs go beyond one billion euros, and are paid for by several countries. Apart from the building costs of the telescopes and satellites, there is also the maintenance expense. In total, taking into account the average life of a large ground or space telescope, each observation hour costs thousands or tens of thousands euros. Therefore, it stands to reason that the use of the telescopes by the first barmy person who promises the moon and stars should be avoided at all costs. Moreover, nowadays, in order to obtain such huge amounts of money, it is better to show a solid image of science, an image that indicates science knows where it is going and has a clear view on problems that have been solved and those that are to be solved when money is available for research and instruments, etc. The image of a pluralistic science, entrenched in discussions about fundamentals, is not sufficiently interesting to attract investment money. Huge amounts of money are not invested for the purpose of allowing the scientist to play at guessing how nature is. Those sums, however, are invested in order to obtain a firm product far away from the speculative wordiness of philosophers. In other words, science is bought, and the one who pays has the right to demand the fruits are superior to those obtained for a lower price. All is accounted for with regard to the sums: the number of publications obtained with a telescope, the number of citations obtained by those papers—called "impact", which is a parameter as related to the quality of a work as is the number of people comprising an audience for various television programs. In the end, informs must talk about how profitable an investment has been. Parameters such as genius, creativity, mental lucidity and other human factors are not included in these informs.

Spontaneity has no place in actual science. Neither, is there a place for fortuitous discoveries favouring those who suspect that something in astrophysics is not following an appropriate path. Almost everything is planned in order according to programming and forecasts made many years in advance (it takes around fifteen years from the beginning of the plans to build a satellite until it is launched). Predictability was seldom present in the long history of science. Many times, science has had to walk back a way, to retrace its steps, before taking some new way or approach. Many surprising discoveries have been done by pure chance. However, contemporary system is apparently surer than the science of any previous time in that there are no historical errors; at least, if there are some errors, then the system tries to delay their discovery as long as possible. It is an apparent paradox that the greater the possibilities of science to observe and make experiments, the greater the obstacles—rather than motivation—for its advancement. Does astronomy move backwards with the advance of technology? There is no doubt that astronomy is an observational science that has larger

"possibilities" with better instruments. However, the control of science by the system is larger when larger telescopes are available, so private initiatives are blocked if they are in disagreement with established standpoints. In this sense, astronomy regresses. That is, the 'possibilities' escalate but the efficient use of these possibilities declines. Telescopes do not 'think' alone and produce casualties when advances in technology result—as is often the case in some sectors of science—in mental atrophy.

**Advance/stagnation of science**

There are many real examples that can be given; many names, and many problems related to astrophysics that were manipulated in favour of a given trend and where the prejudice against alternative approaches can be illustrated. The discussion of these particular theories is not the issue in this paper. Nevertheless, I can speak about some instances of general trends. For instance, the almost exclusive use of gravitation as a basis of understanding many astrophysical problems dealing with the large scale cosmology, galactic dynamics, and formation of large-scale structure, etc. There is a quite strong stream towards this direction. There are alternatives, of course, and we could take, for instance, the case of electromagnetic interactions at large scales, but working with these forces needs a much larger effort than simply working gravitational forces. The uncertainties about magnetic intergalactic fields, for example, are huge. Then, what is to be done? A devil-may-care attitude is predominant. Gravity is used to try to solve these problems and, when magnetic fields are mentioned, pained faces show up as well as expressions indicating "one should not make life difficult for us; we are happy with what we do". However, nature is difficult to understand, and truth may have nothing to do with the positions taken by some scientists who do not want to have disturbed the peaceful tranquillity of their lives. Ockham's razor is usually cited by many who are often pretending to emulate a philosopher. Apparently, Ockham and his razor is the only philosophical reference that many scientists can lay claim to but it is often cited inappropriately because nature's simplicity is confused with the simplicity of what they can calculate. "Nature does not care for analytical difficulties"—said Fresnel in 1826. There are many problems in astrophysics and gravitational interactions are not always the answer. In order to resolve the situation, leaders in scientific research push science toward speculative ways with terms such as super-massive black hole, non-baryonic dark matter, inflation, cosmological constant, gravitational lens which are some times not even understood. Nonetheless, they continue to speak in gravitational terms of reference rather than opening up to, and learning new branches of physics from those to which they have dedicated the past twenty or thirty years. In considering large issues such as the luminosity of quasars, they claim the existence of invisible large black holes, with millions or tens of millions times solar masses. They forget Ockham's razor and it does not matter that they have to use all the patches at their disposal to fill the gaps created by their prejudices. All is possible, except the taking of leave, a departure, from their prejudices. The same thing happens with the topic of intergalactic extinction of which exact knowledge has not been achieved and which, for convenience, is taken as null in all wavelengths up to very large distances. Another example is quasar distances, commonly accepted to be the distance derived from the redshift which is interpreted as being cosmological, in spite of the problems which this interpretation has in explaining certain observed correlations between nearby galaxies and distant quasars. There are other problems that are avoided by looking away, elsewhere, as soon as they are mentioned. Is there insufficient visible matter to justify the kinematics? No problem; dark matter is introduced and everybody is happy. Research advances until it is realized that dark matter cannot be any known matter. Then, another patch is introduced in the established theory and new types of never-before-seen matter are invented: non-baryonic dark matter, which is also useful in solving problems in observing CMBR anisotropies a thousand times lower than expected before the taking of measurements. And inflation is invented, and the cosmological constant is whimsically put or removed in Einstein's equations, according to the fashion, and more and more free parameters are added to a theory in such a way that, if something does not fit the observations, it is a question of changing parameters ad hoc. And when will this fashion of patching the theory, to ensure it accommodates those results (that needed to be ironed out) in order to make a posteriori predictions, be finished? Perhaps, when somebody realizes that there is a failure in the base premise of the piece-by-piece-built construction. Then, it will be the time to throw everything into the dustbin and begin again in some other place. A very clear historical example comes to mind in the Copernican revolution that tore down the highly patched astronomy of Aristotle-Ptolomeus. This is precisely what is presently being pursued simply for avoidance, at all costs.

Maybe all the "alternative" theories are wrong—maybe. Nevertheless, can we be one hundred percent certain that the standard scenarios are correct in order before we reject systematically all the alternative proposals just because they are against an orthodox view? I do not think so. However, the system acts apparently as if it holds the final theory in its hands. The system has a set of modern patched theories, like those of Aristotle-Ptolomeus, and it is afraid of the loss of its privileged status. Galileo had to fight hard against the mainstream in his time, and the passage of history has, in many respects, changed little. In fact, it seems that nothing changes. It is pitiful that nowadays propaganda sells us the idea of freedom that is so far away from the circumstances of four or five centuries ago, yet we really live with the same dogs, although in different collars. At least we have progressed somewhat for certainly nobody is burnt at the stake. At worst, somebody may be exiled from their kingdom and life made impossible for someone in order that they do not publish or otherwise advance in their research. It is also pitiful that an image about cosmology, for example, may be sold such that everything is perfectly clear and only a few parameters remain for high precision fitting. It is probable that the basis on which actual cosmology (a relatively young science, if it can be called a science at all) has been developed is completely incorrect. However, the system continues to build, rebuilding itself ever more quickly over ever increasing quicksand. It seems that nothing has been learnt from history; that the economic interests which power the business
of science are conveyed into thinking that a solid knowledge is firmly treading the 'good way'; into taking that 'way' forward regardless of the possible sabotage and disagreeable, critical elements.

When all these arguments are related to an orthodox scientist, the answer is usually that science is objective and, therefore, a first theory is supported rather than another second theory, because further proof was obtained in favour of the first one, rather than the second one. These words sound very nice, and they even appear honest. However, in the light of all that has been said in the present paper, one must consider that not everything is so honest nor do I say that everything is pure manipulation either. No, there are many cases in which nature shows itself clearly enough in the experiments and observations, and the conclusions are irrefutable. But there are many turbid cases, belonging to turbid sciences such as cosmology, in which the power of manipulation is stronger than nature.

It is true that certain standard theories work better at explaining the data. However, the number of persons involved in a particular theory, and in patching it here and there is not generally told. It is said that the Big Bang theory, for instance, has defeated competing theories. Of course, this theory has thousands or tens of thousands of researchers who in some way are involved or interested in the theory being correct otherwise their work of their lives would be jeopardized. Whereas, an opposing theory may be defended by a small number of researchers who can be counted with the fingers of one hand. Even if they are very good scientists, they cannot compete to produce patches and to spread propaganda on the same scale as do the huge numbers of orthodox researchers. These researchers have to fight against the system without money, without students, without telescope time. Personally, I think that cosmology is not a serious science and I do not believe any theory, neither the Big Bang nor the competing theories. In any case, to place these conditions in the context of a sporting framework, the game is not fair nor does it seems appropriate to talk about defeat when the real issue is abuse.

Objectivity in the scientific method is usually aimed at a target. However, in turbid matters, the method to be applied is usually not very objective and basically is as follows:

- Given a theory A self-called orthodox or standard, and a non-orthodox or non-standard theory B. If the observations achieve what was predicted by the theory A and not by the theory B, this implies a large success to the theory A, something which must be divulged immediately to the all-important mass media. This means that there are no doubts that theory A is the right one. Theory B is wrong; one must forget this theory and, therefore, any further research directed to it must be blocked (putting obstacles in the way of publication, and giving no time for telescopes, etc.).

- If the observations achieve what was predicted by theory B rather than by theory A, this means nothing. Science is very complex and before taking a position we must think further about the matter and make further tests. It is probable that the observer of such had a failure at some point; further observations are needed (and it will be difficult to make further observations because we are not going to allow the use of telescopes to re-test such a stupid theory as theory B). Who knows! Perhaps the observed thing is due to effect "So-and-so", of course; perhaps they have not corrected the data from this effect, about which we know nothing. Everything is so complex. We must be sure before we can say something about which theory is correct. Furthermore, by adding some new aspects in the theory A surely it can also predict the observations, and, since we have an army of theoreticians ready to put in patches and discover new effects, in less than three months we will have a new theory A (albeit with some changes) which will agree the data. In any case, while in troubled waters, and as long as we do not clarify the question, theory A remains. Perhaps, as was said by Halton Arp, the informal saying "to make extraordinary changes one requires extraordinary evidence" really means "to make personally disadvantageous changes no evidence is extraordinary enough".

**Unofficial science**

The system really invites being left alone. I am actually convinced that if somebody wants to make something important—here, again, I remark that this is not only applicable to the sciences but, in general, to any human Mafia with the name of 'culture'—it must be done away from officialdom, and perhaps in free time and laborious study by oneself. The problem for the sciences with this position is thence the precarious or even nullified possibilities available to thus observe or make experiments, not to mention the bad reputation associated with free-thinking occurring away from the official institutions. Since the expenses for the necessary materials are very high, the possibility of doing high-level empirical research from the periphery, in any field, is practically nought. The only possibility is pure theory/speculation, or perhaps feeding of empirical data produced by other scientists which is, in fact, quite frequent.

I receive very frequently—nearly every one or two months—by e-mail new theories from amateurs who try to throw down all the well-established physics to leave space for new and often ridiculous theories, or cases of cosmological theories that are failures in even the most basic aspects. This kind of work has practically no reference in professional journals, and tends to cite popular books on science. Rather than studying a particular scientific problem, they talk about very general matters. For instance, they try to throw down all the known physics. Precisely because of that, the independent research carried out away from official institutions finds problems of credulity; for each researcher with enough preparation who wants to do serious things, there are thousands of 'barmies' on the planet who dream of creating a theory of physics inspired by the heavens such as poetry, which demolishes all the past and opens a new era in the history of science. I once heard on a radio program an interview with a carpenter who had never studied physics, but had just read some popular books on physics, without trying to understand anything about maths. The carpenter said that he had written seven books about black holes, and he complained that he could not publish any one of them. I do not want to judge negatively the efforts of some amateurs, who perhaps have read some popular-science book by Hawking and think that they are able to work as researchers. I do not want to act as a part of the system that castrates any attempt at originality just because it is challenging. Nevertheless, the reality is that amateur's theories have a lot of failures and inconsistencies because they have no knowledge independent of the ideology. And the result is that the thousands of 'barmies' in the penumbra do not get to listen to the voice of some possible genius who could be in their midst. Therefore, autonomous research activities do not have a high credibility, and one must use official mechanisms in order to be listened to by other professional researchers.

**Attitude of philosophers to science**

It is probable that some will identify the present manifesto as a philosophical criticism, a charge made by many against science. I think that the present way of thinking is philosophical. However, it must not be confused with the types of presentations made by the self-claimed professional philosophers i.e., those who have an academic degree.

Indeed, it is not often one finds this type of criticism about science. There is criticism, of course, but very detached from contact with the problem and often no more than a paraphrasing of metaphysical speculation that has very little to do with the above mentioned problems. The philosophy of science as nowadays taught in the faculties of philosophy is, indeed, a philosophy of anti-science. It is taught that scientists are inept and do not know how to think while the professional philosophers are those who are able to give sense and meaning to science. There are several approaches to the philosophy of science that I will not be discussing because that is not the goal of this paper. There is a wide range of positions that could be taken. For example, the openly anti-scientific position that compares science with religion and holds the view that there is not any truth in scientific knowledge or that the science of an African tribe's witch doctor is comparable to western science. Then there are the less crazy ones, limited to explaining to the world—in very thick volumes—what a hypothesis is, or the falsability of a theory, and those trivialities which are well known to any scientist since early education, and which do not reveal anything not already known. Apart from these efforts, professional philosophers make very few attempts to understand the present-day problems of science and perhaps, some of these problems are only mentioned in order to discredit science in general. Sentence such as "this agrees what we had said..." and trying to sell some of the metaphysical and paranormal (in Spanish "para anormal" means "for an abnormal person") creeds that are usual merchandise of modern sophists. A Spanish proverb says: "under the heavens, everybody lives on one's work".

Why is this kind of criticism so infrequent among the works of professional philosophers? I think there are two main reasons: 1) they do not have knowledge of the science from close quarters but through reading books which do not reflect the real problems; 2) they are not interested in revealing the problems of another profession because they themselves share the same problems in even greater magnitude. Obstruction of the freedom to initiate a research line or ideology is more prevalent in the faculty of philosophy than in science. Philosophy congresses are simply imitations of scientific congresses. Censorship of publications is more evident (most being confined to local town dissemination rather than international); they have practically no objective criteria and there are no empirical data, so a paper can be rejected whimsically, without even producing information as to the cause or reason for the rejection. Work positions are nearly always handpicked. Communication with the press, or promotion for the publication of books by editorialising is in the service of the corresponding supervedettes. Propaganda decides the survival of philosophical nonsense, etc. In this panorama, what has the 'office-philosopher' to say about science? Therefore, it is not strange that there should be silence about these aspects. They prefer to dedicate their efforts to ascertaining what is the meaning of 'truth' or how many types of reason exist or many exercises of language analysis or the classification of the different schools with different "-isms". As has been said, under the heavens, everybody lives on one's work. The problems of science are not going to go away, nor are they to be resolved by any paid philosopher. These questions are things to be discussed by scientists themselves, and from the inside looking out.

**Some final optimistic notes**

In short, I see with certain pessimism the actual state of astrophysics, as well as of other sciences with similar problems. All the circumstances above described may lead one to conclude, that the actual 'product' from the branch of science known as astrophysics has become prostituted in many senses. According to the dictionary, one of the meanings of 'prostitution' is the use of talent or ability in a base and unworthy way, usually for money. This is what astrophysics and world's oldest profession have in common. These problems are reflected in many other fields of culture, as well as in our own society. Everything produced contaminates everything else. There are no isolated problems; any human activity is a reflection of the environment that surrounds it. We live in a rotten society that deceives itself. What else could be expected from science in such a society?

I recognize that my criticism is not objective. In contrast with the exaggerated optimism of other, perhaps my pessimism is exaggerated too. Perhaps my view is somewhat disproportionate. Well, each can be judged. I am simply expressing my opinion and everybody has an opinion. This is not a pamphlet with political or sectarian ambitions. I am not interested in convincing anyone of any claim. I do not think that this text is useful for trade-union claims that demand the rights of science workers. Rights are not the issue here but facts are: to know how nature is. Making high quality science is the issue here. It is not helpful to claim some "right", because the present problems of scientific research will not be solved with the increase of bureaucracy; they would simply worsen. Neither is it a matter of asking for further money to solve the existing problems. Quite the opposite: the more money is invested, the more the system becomes a Mafia. I am pessimistic about even finding politically correct solutions in the actual social context.

Anyway, I do not want to finish this text without arguing that everything is not black and although not totally satisfying, there are certainly reasons to have some degree of optimism. Truly, in spite of the problems that exist in the scientific infrastructure, I think that science can be made and there is a net advance. It is not the huge advance claimed by propaganda, but there is some advance. It is slow, with many errors that are very slowly being corrected, but there is some advance and our knowledge about the universe is maturing. In other epochs, there were also many difficulties that were barely overcome. However, it seems that there is a historical mechanism that with the independence of human interests, polishes and filters the most solid of knowledge as time goes by. Probably, it is because the created interests vanish gradually with the advance of generations, and it is only after some tens or perhaps hundreds of years, that ideas with intrinsic value are distilled and survive to present us with their wisdom. Indeed, history is not always fair. Many good ideas are forgotten and are not recovered until they again rise to the surface of independent thought. Copernicus had to rediscover what Aristarchus of Samos knew seventeen centuries before. There are many cases of historically famous researchers who have stolen merit from persons unknown. Neither is history itself perfect for, after all, it is also human. Nevertheless, I believe that there is something great in astronomy, in physics, in all the natural sciences that allows the human being to look beyond its present place and to arrive at some understanding of what goes on beyond the insignificant meanness of spirit that so often pervades our existence. There is a Nature; there is a Cosmos; and we walk towards the understanding of it all. Is it not wonderful? There are many charms in the profession; as many charms as in love provided, of course, that they are not in the service of mercantile aims.

# ¿Qué tienen en común la astrofísica y la profesión más vieja del mundo?[4]

Martín López-Corredoira

Instituto de Astrofísica de Canarias

C/.Vía Láctea, s/n

E-mail: martinlc@iac.es

URL: http://www.iac.es/galeria/martinlc/

Dedicado a Eduardo Simonneau,

maestro del desengaño

La vida es la mejor maestra, mucho mejor que la Universidad. He aprendido algunas cosas interesantes sobre astrofísica durante mis últimos 10 años en la investigación, pero no sólo de los astros he aprendido cosas. En la Tierra he encontrado cosas interesantes que merece la pena aprender. Intentar entender la mecánica de las estrellas, galaxias, etc. es bello y me alegro de dedicar mi tiempo a tan noble profesión. Sin embargo, uno debiera tener siempre en mente que estamos en la Tierra, rodeados por otros hombres, y a ese respecto debemos preocuparnos del suelo que pisamos, y no sólo mirar al cielo.

No creo que la astrofísica sea un caso especial dentro de las ciencias. Tampoco creo que las ciencias sean un caso especial dentro del mundo de la administración de la cultura. Todos navegamos en el mismo barco: el mundo en la actual era capitalista. Sin embargo, me concentraré en la investigación de astrofísica porque conozco mejor este mundo (también conozco algo de las facultades de filosofía, otras casas de ... que probablemente tengan más que criticar que los institutos de investigación científicos, pero no es el tema principal aquí). En el presente artículo, quiero narrar mis impresiones de este mundo que he conocido de cerca de un modo abierto y sin autocensura, diciendo las cosas como me parece que son, y sin preocuparme de si esto es simpático al que lo lea o no, o de si me publicarán esto en algún lugar. Creo que el único método de alcanzar la verdad es no temer nunca decir la verdad y anteponer ésta a otros intereses, como puede ser el medrar dentro del sistema, conseguir un puesto, publicar en revistas de prestigio, etc. Probablemente, dado que se trata de apreciaciones subjetivas, puede que haya errores en lo que se dice aquí, pero no importa mientras la dicha sea buena: decir honestamente qué pensamos de las cosas.

---

[4] Publicado en: López Corredoira M., Castro Perelman C. (Editores), 2012, A contracorriente: Un análisis crítico por científicos sobre cómo se hacen la física y la astronomía, Common Ground Publishing España, Madrid, pp. 119-144.

El funcionamiento de las actuales instituciones científicas es algo complejo. Uno sólo comienza a entender la mecánica social de éstas cuando lleva un tiempo trabajando en ellas, desde dentro. No creo que un filósofo o un sociólogo que lea cuatro libros y no conozca en directo lo que se cuece en los centros de investigación llegue a inferir correctamente una teoría plausible de mecánica social de las ciencias. Si acaso, podrá inferirlo por comparación y extrapolación de conductas en otras instituciones, pero no leyendo unos libros dado que no hay practicamente publicaciones que reflejen la situación real. La verdad no siempre está al alcance de las ratas de biblioteca, porque hay verdades que no se escriben o su difusión es harto limitada (como le puede corresponder a este mismo texto), quizás porque no interesa que se conozcan.

**Estudiantes**

El primer contacto con la investigación lo tiene uno cuando prepara la tesis doctoral. Aquí, como en muchas otras disciplinas, el sistema adopta una posición clara: "las cosas son como nosotros decimos; o lo tomas o lo dejas". Si uno quiere trabajar en la investigación debe ponerse al servicio de un programa predeterminado por las autoridades del sistema. El papel del estudiante, si quiere conseguir apoyo económico, y si quiere en general tener un mínimo apoyo en los departamentos, es obedecer y asimilar la tradición del departamento en el que trabaje.

En tono entre bromista-irónico y cínico, con cierta presunción no muy desencaminada, se suelen llamar a los estudiantes "esclavos". Ellos son los encargados de realizar la mayor parte de las tareas monótonas de la investigación (observación durante largos periodos de tiempo en los telescopios, reducción de los datos, etc.) al servicio del grupo en el que trabaja. Existen también algunas figuras inferiores en la jerarquía del sistema: los becarios temporales (por unos pocos meses), que en muchas ocasiones son alumnos que aún no han terminado la carrera y por lo tanto están por debajo de los doctorandos. A estos se les suele llamar "esclavos de verano", porque su contrato es por los meses de verano, y en ese período no da tiempo a que aprendan nada de la investigación; con lo cual se les utilizaba como mano de obra barata: en pocos días se les enseña una tarea mecánica y luego se los deja todo el verano realizando esa misión rutinaria.

Esta apreciación del trato a los estudiantes no es realmente aplicable a todos los casos. En mi caso, por ejemplo, no lo fue. Sin embargo, me consta que esta realidad explotadora está bien extendida y es bastante más común de lo que uno desearía. Me atengo a las conversaciones directas mantenidas con distintos investigadores más que a datos estadísticos publicados por algún organismo oficial, claro está, pero juzgo que las fuentes son suficientemente representativas.

Hay casos en que el estudiante, realizando la mayor parte del trabajo, no puede escribir los resultados que derivan en las consiguientes publicaciones científicas, sino que eso lo realizan directamente los jefes, poniéndose a sí mismos de primeros autores. Al estudiante se le dice que todavía no sabe lo suficiente como para escribir lo que verdaderamente es su trabajo. Hay casos en que el director de tesis deja al estudiante colgado cuando ve que las cosas no funcionan como quiere. Hay casos en que el director roba ideas al estudiante. Hay casos en que al estudiante se le ha acabado la beca, porque en los 3 o 4 años que dura la misma lo han tenido explotado con otras actividades al margen de su trabajo de tesis, o su jefe no ha tenido tiempo para atender a las explicaciones de lo que el estudiante produce, y tiene el estudiante que arreglárselas para sobrevivir al tiempo que termina su trabajo.

Pocos son los casos de jefes que se sientan con el estudiante a hacer conjuntamente un trabajo. Lo normal es que los primeros días pierdan un tiempo con él para explicarle cómo se hacen las cosas, y luego es el estudiante el que debe realizar las tareas rutinarias. El jefe se dedica a dar ideas, si las tiene; y si no las tiene simplemente se dedica a hacer correcciones menores. El estudiante emplea semanas y semanas delante del ordenador luchando con un programa de ordenador que no sale como uno desea, unos cálculos engorrosos o simulaciones que llevan mucho tiempo. Pasa noches enteras pegadas al telescopio (el jefe suele ir pero sólo la primera vez para explicarle al pupilo como funciona la máquina, o cuando se trata de alguna novedad o de unas observaciones extraordinarias fuera de la rutina; en cualquier caso, es muy normal que se vaya a medianoche a dormir y deje al estudiante hasta el amanecer con la rutina si no hay nada que le retenga), semanas enteras en el telescopio, para luego bajar con sus cintas de varios gigabytes de datos y "reducirlas", es decir, procesarlas, extraer información de las observaciones. Esta labor requiere normalmente varios meses, y si las cosas se complican por algún error en el procedimiento y hay que repetir el proceso pues lleva más tiempo todavía.

Mientras, el jefe dirige, pone ideas. Dice: "Está bien, pero podrías hacer esto, y lo otro, y lo de más allá". Esto le llevará al estudiante fulanito una semana hacerlo. Lo otro le llevará dos semanas y al final se dará cuenta de que es inviable. Lo de más allá seguro que es una tontería pero el jefe no se convence hasta que se comprueba con algunos cálculos más (que, por supuesto, hace el estudiante). Al final le entrega los resultados al jefe, y luego éste dice: está bien pero me gustaba más como estaba antes.

¿Cuál es la objeción que se le puede poner a esta situación? ¿Acaso no es normal que el maestro enseñe al alumno, y el alumno haga lo que se le diga mientras está aprendiendo? Cierto, así debe ser. Lo cierto es que el alumno recién licenciado no conoce mucho del campo específico sobre el que ha de trabajar y debe ponerse al día. Pero no es tampoco un novato sin conocimientos. Suele realmente tener mayores conocimientos generales de la astrofísica en conjunto que el especialista que sabe mucho pero sólo de su área. Además, el alumno posee algunas ventajas sobre el maestro en este caso: es más creativo, tiene la mente más libre de prejuicios y puede aportar nuevos y frescos puntos de vista a la investigación que desarrolla en vez de seguir una tradición anquilosada en los intereses de quien ha invertido toda su vida en una idea. Un doctorando puede tener ideas si le dejan tenerlas, incluso fuera del carril predeterminado para él. Al respecto, parecería más lógico que el trabajo monótono esté en manos de quien tiene la creatividad agotada, de las viejas autoridades de la materia que ya no van a producir más que copias de lo que siempre han hecho. Sin embargo, el sistema de la ciencia no sigue esa lógica sino la del poder: donde hay capitán no manda marinero. Por ello se suele cargar con los trabajos más rutinarios a los estudiantes; no para que aprendan (pues aprender se aprende la primera vez, pero no haciendo cien veces lo mismo), sino para que produzcan. Hay algunos pocos casos, sí, de doctorandos que investigan por cuenta ajena al margen de sus obligaciones con el director de tesis, y al margen del campo en que se circunscribe su tesis (eso sería lo que yo recomendaría a los futuros estudiantes), pero eso no es lo que hay generalmente.

Me han contado una vez una anécdota al respecto, que no sé si es realmente cierta, pero tiene toda la pinta de que sí: un estudiante habla con su director y le dice "he tenido una idea", y éste le contesta "¡Ah!, pero ¿tienes tiempo para pensar?". Así es como se forma a los futuros científicos, haciéndoles pasar el tiempo con mil tareas rutinarias que no les dejen tiempo para pensar, al menos para pensar por libre. Sí se incentiva y se alaba a quien piensa cómo corroborar una y otra vez la idea de una autoridad de la ciencia, pero tener tiempo para pensar y hacerlo por cuenta propia sin que nadie haya dado permiso es algo que el sistema realmente no incentiva. Al contrario, se desalienta la iniciativa propia bajo el peso de razones tales como que lo que está establecido está bien establecido. Se preparan trabajadores de la ciencia, currantes, no pensadores.

¿Cuál es la objeción que se le puede poner a esta situación?, sigo preguntando. Principalmente, que no se da una formación creativa sino más bien industrial (de cómo producir chorizos en serie), y también que se agota el máximo periodo creativo del científico entre estos avatares. Hemos de tener en cuenta que, a lo largo de la historia de la ciencia, la mayor parte de las grandes ideas han venido de mano de jóvenes. Si se utilizan como esclavos, o mejor dicho obreros de la ciencia, a los jóvenes que pueden aportar nuevas ideas, el resultado será la perpetuación de lo antiguo y el estancamiento del intelecto, aunque, aparentemente, si uno hiciera caso a los medios de comunicación, una revolución científica se está produciendo todos los días.

**Postdocs, plazas fijas**

El investigador que quiere vivir del mundillo de la investigación debe aspirar a una plaza fija, es decir, llegar a formar parte del cuerpo de funcionarios del estado, sobradamente conocido por todos por su eficiencia a todos los niveles. Bromas aparte, el caso es que la vida y las motivaciones del investigador postdoctoral está marcada y orientada con el fin último de la consecución de la plaza. En el mejor de los casos, el funcionario seguirá estando motivado después de alcanzado ese objetivo, pero en muchos otros casos ocurre más bien lo contrario.

Un estudiante con la tesis recién leída rara vez obtiene inmediatamente la plaza fija, sino que ha de pasar antes por varias instituciones (necesariamente alguna en el entranjero, aunque hay excepciones) con contratos temporales conocidos como "postdocs". Esto, realmente, me parece de lo más inteligente que tiene el sistema, porque al menos el investigador dispone de unos años de su vida para buscar sus propios caminos en la investigación, y al mismo tiempo se impide el estancamiento temprano que suelen producir las plazas fijas concedidas tempranamente.

El status "postdoc" constituye una posición jerárquica de obrero de la ciencia por encima del doctorando, pero por debajo del investigador de plaza fija. También sobre estas figuras recae una buena parte de las tareas rutinarias de la ciencia, pero en menor medida que los doctorandos ya que, en bastantes casos, disponen de movilidad propia; en muchos otros casos, por el contrario, son obreros contratados para un programa predeterminado. El hecho de que se vaya al extranjero o no me parece poco relevante, a no ser para ganar soltura con el inglés u otros idiomas, porque hoy en día la investigación está bastante globalizada y poco puede aprender uno en otro país que no pueda verse en el de uno propio. En todos los sitios cuentan lo mismo y hacen lo mismo, con diferencias menores. Influye quizás más el tipo de investigación con el que uno toma contacto que el hecho de irse al extranjero. Por lo general el investigador sigue rodeado del mismo tipo de entorno que le acompañaba en la tesis con lo cual tampoco aprende muchas cosas nuevas.

Hay que decir que no todos los doctores siguen su carrera como investigadores. En muchos casos las motivaciones de la vida privada impiden la movilidad requerida en la profesión o bien no se puede optar a ella debido a la alta competencia existente, que posibilita que algunos, no todos, puedan continuar. Para poder ganar una plaza postdoc es necesario tener un buen curriculum, que no tiene por qué ir asociado con la genialidad sino más bien con la capacidad para currar, y el apoyo o recomendación de alguien de dentro del sistema. Sin la recomendación adecuada la carrera se puede truncar, con lo cual se hace necesario buscarse simpatías en el mundillo (y una manera de buscar simpatías es seguirle la corriente a las líneas generales de investigación sin intentar ni siquiera hacer un frente crítico). Con respecto al curriculum, lo normal es que el peso mayor venga dado por el número de publicaciones en revistas especializadas. Digo número (cantidad), no calidad, porque realmente el parámetro predominante va a estar más relacionado con lo primero. Se valora la calidad cuando la comisión que evalúa a quien concede la plaza es especialista en ese mismo campo y del mismo bando (es decir, que no es de la competencia con otras teorías para explicar lo mismo). Dado que cada especialista piensa que su campo es el más relevante siempre encontrará apoyo el curriculum en cuanto a calidad cuando está orientado a los intereses del tribunal que lo juzga. De lo contrario, será un simple número: al peso, como en los trabajos de la escuela.

Esta manera de evaluar el esfuerzo del investigador será una constante a lo largo de la vida del mismo; ya sea para conseguir una plaza, o para conseguir tiempo de telescopio, o para conseguir dinero para su proyecto, etc. Más adelante me detendré en estos aspectos. En lo que concierne a las posiciones postdoc, hay que ver en este sistema de evaluación una presión indirecta sobre la supuesta libre elección de caminos en la investigación para enfocarlos en la medida de lo posible hacia los ya dados; tanto por el número de publicaciones (que para ser elevado debe ser poco crítico), como por las simpatías que puedan adherirse a corto plazo. Ello será lo que le permita acceder a otras postdocs, o a la plaza fija, si es que antes no abandona u opta por buscarse la vida de otro modo.

En mi experiencia, por ejemplo, aunque no he tenido demasiados encontronazos, he estado trabajando en algún campo no muy ortodoxo y he podido por ello comprobar los problemas que se originan cuando uno trabaja en lo que no le mandan. Debido a mis acercamientos a investigadores que trabajan en teorías científicas poco ortodoxas, para discutir sobre ciertos datos o intercambiar opiniones diversas, he tenido que escuchar más de una voz que me aconsejaba, por mi bien, alejarme de estos campos de investigación y de toda posible relación con estos investigadores, pues el resto de la comunidad podría llegar a relacionarme con ellos y eso sería una traba para conseguir una posición postdoc o plaza fija para el futuro. Cuando fui invitado a dar una charla sobre el tema, un investigador senior me dijo que dar esa charla supondría olvidarme de conseguir cualquier tipo de plaza allí donde la iba a dar. Esto no es un chantaje, pero realmente se le parece.

De mi propia experiencia y de otras que conozco deduzco que para tener las puertas abiertas uno debe ser servil y poco crítico. Debe producir mucho, pero sin grandes aspiraciones de querer decir algo importante. Es triste tener que decir que las plazas que he conseguido me las han dado por los trabajos que yo considero menos interesantes, y que aquellos que yo considero interesantes no me han dado más que problemas, discusiones, dolores de cabeza y falta de atención. Es triste, pero así es, y me consta que no soy un caso aislado.

## Publicaciones, árbitros

Los frutos de la actividad científica se recogen en las publicaciones especializadas, en las revistas que luego serán leídas por otros especialistas y distribuidas por todas las bibliotecas del mundo de los institutos de investigación (a precios desorbitados, bien para publicar o para comprar las revistas, que sólo las instituciones adineradas pueden pagar, claro; el negocio de las revistas científicas no es una cuestión baladí aquí). Las revistas son hoy en día una herramienta de comunicación potente: escritas en un lenguaje internacional, el inglés, y con una accesibilidad envidiable. Hay que reconocer que el actual sistema de publicaciones científicas es muy superior al de cualquier otra área de cultura, donde falta la unificación del lenguaje y las publicaciones están dispersas en múltiples revistas locales de difícil acceso.

Como es bien sabido, el control de las comunicaciones y el ejercicio del poder están íntimamente ligados. Creo que no descubro nada nuevo con tal afirmación. Es por ello que el sistema, lejos de permitir la libre publicación entre los profesionales de sus resultados, funciona con un sistema de censura teóricamente ideado como control de calidad pero cuyas funciones se extienden frecuentemente a un control del poder. Quien quiera publicar en estas revistas ha de someterse al dictamen de un árbitro escogido por los editores de las revistas, que diga si ese artículo se acepta o no. El hecho de que el árbitro sea la mayoría de las veces anónimo, porque él mismo así lo desea, ya indica de por sí que su actividad no es siempre honesta. He tenido incluso un caso en el que el editor era anónimo y sólo conocíamos el nombre de la secretaria. Si fuese honesta esta actitud no se escondería detrás de un anonimato, quizás por temor a que se le señale como detractor. Quien cree realizar una buena labor de consejo para una revista no tiene por qué esconder su nombre.

Por lo general, los artículos se envían a árbitros que son especialistas en la materia y que pueden aportar sus conocimientos para mejorar la calidad del artículo a publicar, o detectar errores de cálculo si los hubiera, o contradicciones con algunos datos, etc. La idea en principio es buena, y lo sería si el proceso de arbitraje fuera siempre objetivo y desinteresado. Creo que no es ése el caso. Hay una buena cantidad de ocasiones en las que los conflictos de intereses dictaminan el destino de un artículo antes que otra cosa.

De mi experiencia publicando artículos científicos en revistas internacionales con árbitro he observado que los informes de los árbitros rara vez detectan errores en los cálculos o procesos de reducción de datos, pues no tienen paciencia para realizarlos de nuevo ni chequear los algoritmos, y, aparte de los detalles menores (cambiar una gráfica para que se vea mejor, explicar algún párrafo mejor, citar algún otro artículo [en muchas ocasiones el árbitro aconseja que se le cite a él y a sus colaboradores], etc.), las objeciones mayores tienen casi siempre que ver con la opinión o el convencimiento que el árbitro tenga sobre el contenido de lo que se va a publicar. Generalmente, según mi experiencia y otras que he podido escuchar, cuanto más controvertido es el tema que se toca, más desafiante a las ideas establecidas y más novedoso su enfoque, más problemas se encuentran a la hora de publicar, teniendo muchas posibilidades de que se rechace la publicación. Por el contrario, cuando uno escribe un artículo que hace y dice lo mismo que otros cientos de artículos ya publicados sobre el tema (cambiando quizás algún parámetro si se trata de un modelo teórico o cambiando el objeto observado dentro de un mismo tipo con respecto a otros trabajos que hay en la literatura o haciendo lo mismo con datos de mayor calidad), y que llega a las conclusiones que todo el mundo sabía y sobre el que todo el mundo (especialmente el propio árbitro, que suele ser un representante de las ideas ortodoxas) está de acuerdo salvo en puntos secundarios, entonces se encuentra uno en general con una actitud menos beligerante por parte del árbitro, incluso recibe uno felicitaciones.

El problema de fondo es el siguiente: los árbitros son personas que han dedicado toda su vida a investigar unos pocos problemas de un área. Son personas ampliamente reconocidas en su campo y deben sus status social a sus aportaciones dentro del campo. Como personas de experiencia y de prestigio, que a veces se traduce en un exceso de vanidad, suelen pensar algo como lo siguiente: "Yo soy un gran especialista en este campo. Conozco las ideas interesantes y cruciales sobre el mismo. Si se presentase una idea nueva, unas de tres: o es poco interesante, o es errónea, o si no ya se me hubiese ocurrido a mí. Entonces, si alguien me presenta un nuevo trabajo que pretenda tocar temas cruciales, o es una continuación de mi propio trabajo y mis propias ideas y aquellas que yo he absorbido, o es erróneo". Además, está la propia competencia que supone que alguien publique una teoría o interpretaciones diferentes a las que el árbitro sostiene. Quizás sea exagerado atribuir este pensamiento a muchas "autoridades de la materia", pero creo que en menor o mayor grado algo de esto se da; aunque, por supuesto, si se le preguntase al vanidoso negaría rotundamente esta vanidad. Lo cierto es que este mecanismo psicológico, aunque no explícito, puede darse en la mayor parte de los casos donde se discute sobre lo convincente o creíble de una teoría, o sobre cualquier otro enfoque subjetivo. Cierto es que la ciencia tiene un contenido objetivo, y los datos y las matemáticas están ahí, independientemente de lo que se opine sobre ellos, pero la interpretación de los datos y la plausibilidad de las teorías es algo que está muy sujeto al factor humano, a las creencias en muchos casos, a los prejuicios, y esto toma mucho peso en la censura de las publicaciones científicas. Bueno, también hay que decir que hay muchos árbitros que realizan estupendamente su labor.

¿Cuál es la consecuencia de esto? Tiene un efecto positivo, sí: no dejar publicar centenares o miles de artículos erróneos con ideas disparatadas que no van a ningún lado; sin embargo, el lado negativo es también obvio: el entorpecimiento del pensamiento y las pocas ideas interesantes que pudiera haber. Se premia el trabajo bien hecho pero sin ideas, un trabajo de artesano especialista. Se condena la creatividad. El sistema parece dar a entender que no necesita ideas, parece estar convencido con su conjunto de máximas autoridades de que a la astrofísica o cualquier otra ciencia sólo le falta trabajar algunos detalles. Se da por sentado que la base de lo que se conoce es correcta, que las teorías actuales son más o menos correctas y que sólo se necesita mano de obra para ajustar algunos parámetros o cosas de menor importancia. Una revolución copernicana es totalmente impensable dentro del actual sistema, aun cuando la verdad fuese otra de la actual. A tal respecto no hay muchas diferencias entre la actual academia y la Universidad de los s. XVI o XVII, ceñida a la iglesia y los textos aristotélicos. No es que la ciencia sea como la religión de antaño, como se la ha acusado muchas veces. Los argumentos científicos son bien distintos de los religiosos. Sin embargo, los procedimientos de los grupos humanos que presumen de tener la verdad en su manos guardan muchas semejanzas.

Una excepción importante a esta censura es la existencia de los "preprints" electrónicos "astro-ph" (u otros nombres para otros campos de la física o las matemáticas). Éste es, en mi opinión, el esfuerzo más importante para abrir las puertas a la investigación en el cual el propio autor puede poner su artículo sin el control de un árbitro[5]. Normalmente, los artículos son puestos en astro-ph una

---

[5] P. D.: Lamentablemente, a comienzos del año 2004, dos meses después de que pusiera este artículo en arXiv.org en la sección de astro-ph, se ha impuesto una nueva política que restringe la libertad de difusión de ideas. El nuevo sistema requiere que, si alguien que nunca ha colocado un artículo en una sección de los archivos quiere hacerlo, debe conseguir el permiso de un científico que lo usa regularmente. Incluso alguien que use frecuentemente una sección como astro-ph o quant-ph no es libre de colocar artículos en otras secciones a menos que consiga un permiso de un colega después de que revise dicho artículo. En los últimos años, el sistema de censura se está volviendo más refinado: se suprimen los nombres de aquellos individuos que dan consentimiento para colocar artículos que no son bienvenidos por la ciencia dominante de la lista de posibles colegas revisores. En mi caso, después de dar soporte a algunas personas que querían publicar algunas ideas

vez han sido aceptados en una revista con árbitro de prestigio, pero el autor también puede poner el artículo en astro-ph antes de ser aceptado por una revista. También hay algunas revistas impresas sin censura, pero éstas son revistas pequeñas que prácticamente no se leen. Sólo astro-ph tiene una difusión elevada. La contrapartida de esta libertad de prensa en los preprints astro-ph es que los artículos no son reconocidos oficialmente hasta que son aceptados en una revista de prestigio. No pueden ser usados para apoyar ninguna petición de tiempo para telescopios, por ejemplo. No pueden ser considerados como artículos para argumentar ideas contra otros enfoques. Pueden incluso ser ignorados como si no existieran; si alguien pregunta a una autoridad de la materia sobre un artículo no aceptado en astro-ph, puede contestar sencillamente que el artículo no está publicado y, por lo tanto, puede ignorar sus contenidos como si no existiera (esto sucede también con muchos artículos aceptados). No obstante, en mi opinión, los preprints astro-ph son una buena herramienta para la investigación. Al menos, algún otro investigador sin prejuicios puede leer el artículo y juzgar por sí mismo su calidad. Es útil también para el autor: en mi caso, he encontrado importantes errores en algún artículo gracias a los comentarios de alguien que lo ha leído en astro-ph; errores que ningún árbitro pudo encontrar, pues están habituados a leer las cosas por encima y rechazar lo que no les cae simpático con un simple desaire sin buenos argumentos. Por supuesto, hay un montón de basura en astro-ph pero también la hay entre los artículos aceptados en revistas de prestigio. Y también puede alguien robar ideas, pero eso ocurre tanto si está publicado como si no. Conozco casos de plagio de artículos aceptados que pasaron desapercibidos en su día y que años más tarde un autor de prestigio reescribe las ideas como suyas, tomándolas como propias. ¡Cuántos autores de la antigua Unión Soviética han descubierto cosas interesantísimas, que el mundo no conoció hasta que algún listo norteamericano cargado de dólares reinventó la pólvora![6]

Otro problema radica en el número de artículos. Sólo en la especialidad de astrofísica, se publican unos 30 mil artículos anuales, cantidad totalmente inabarcable aun por el más asiduo de los lectores. Incluso, dentro de un área bastante restringida, como pudiera ser el estudio de los cometas o las galaxias Seyfert u otros, encuentra uno 500 o 1.000 artículos anuales que tienen relación con el

---

desafiantes (pensaba que estas ideas eran incorrectas, pero que merecían ser publicadas), se me informó que no podía ser revisor de ningún artículo más.

[6] P.D.: Puedo dar un ejemplo reciente con la investigación que desarrollé en Tenerife (España) dentro del equipo TMGS. Entre los años 1994 y 2003, nuestro grupo ha estado publicando algunos artículos sobre la existencia de una barra larga en nuestra Galaxia con ciertas características [ver, por ejemplo, Hammersley et al. (2000, Mon. Not. R. Astron. Soc. 317, L45) o López-Corredoira et al. (2001, Astron. Astrophys. 373, 139)]. En el año 2005, un grupo de astrónomos de EE.UU. asociados con la explotación de los datos del satélite "Spitzer" publicó un artículo sobre el descubrimiento de la misma barra con las mismas características: Benjamin et al. (2005, Astrophys. J. 630, L149). Además, publicaron una nota de prensa con el título "Galactic survey reveals a new look for the Milky Way" (cartografiado galáctico revela una nueva imagen de la Vía Láctea) que fue divulgada en muchos medios de comunicación. ¿Una nueva imagen? ¡No!... esto ya lo habíamos propuesto nosotros años atrás. Benjamin et al. no citan nuestros trabajos cuando hablan sobre la barra. Según algunas informaciones que nos han llegado, nos citaban en una primera versión de su artículo (por lo tanto nos conocían; no es una cuestión de falta de información) pero decidieron quitar esa cita y hablar de su descubrimiento como si fuese algo nuevo para ahorrar espacio.

tema, cantidad aún inmensa aunque sólo sea para ojear por encima. Este número ha crecido y continúa creciendo con los años de una forma incontrolada. Decía en tono irónico Chandrasekhar, uno de los antiguos editores de la revista "Astrophysical Journal", cuando se percató del crecimiento desbordado del número de artículos anuales años después de que hubiera dejado de ser editor: la velocidad de crecimiento del número de artículos supera la velocidad de la luz, pero no hay por qué preocuparse, no hay violación de ley física alguna porque estos artículos no transportan información.

Dado que la inmensa mayoría de estos artículos son "paja", es decir, que no aportan nada importante al campo sino simples detalles prescindibles, las posibilidades de los pocos artículos importantes que pasen el sistema de censura de llegar a tener un impacto en la comunidad son bastante reducidas. Esto significa que, una vez pasada la traba de la censura directa de las revistas, el investigador que se aventure a introducir nuevas ideas en el sistema tendrá que vérselas con una censura indirecta: la superproducción de artículos que apantalla lo que no interesa al sistema. La propaganda es el elemento clave para dar a conocer un artículo, y en esto nuevamente tienen ventaja las "autoridades" de la materia, dado que controlan la mayor parte de los hilos que mueven la maquinaria publicitaria: ellos son los que tienen los contactos adecuados, los que escriben los artículos de revisión ("review", es decir, resúmenes de los descubrimientos científicos en un cierto campo), los que organizan congresos y dan charlas invitadas. Además, la reproducción de ideas estándar tiene de por sí mucha más aceptación porque son muchos los intereses de quienes trabajan en lo mismo; mientras que la difusión de ideas nuevas sólo interesa a los que las crean.

## Congresos

El fenómeno de los congresos, simposia, workshops, escuelas, reuniones o cualquier forma de juntar varios profesionales para que se cuenten los unos a los otros lo que hacen es algo bastante extendido, no sólo en las ciencias sino en cualquier ámbito profesional. Aquí, nuevamente, el número se ha disparado en los últimos años. Allá por las primeras decadas del s. XX, cuando cada mes se descubrían cosas de enorme importancia para el desarrollo de la física (pensemos, por ejemplo, en la relatividad y la física cuántica) se celebraba un congreso de pascuas en ramos, y los congresos internacionales importantes tenían una periodicidad anual o menor aún. Hoy en día, sólo en astrofísica, que es una de las múltiples ramas del conjunto de todas las investigaciones de ciencias físicas, se celebran en torno a 200 o 300 congresos internacionales anuales, pequeñas reuniones locales o nacionales aparte. Lo más triste del asunto es que el desarrollo conceptual de la física de hoy en día está muy por debajo de lo que se alcanzó a principios del s. XX.

Las vacaciones pueden ser una razón para ir a los congresos. Muchos de ellos se celebran en lugares exóticos o turísticos, que permiten a los líderes de la ciencia y a sus amigos disfrutar unas vacaciones a cargo de los fondos públicos. El fin principal de los congresos sin embargo no está en promover el turismo sino en la difusión de la información en un microcampo de la astrofísica (u otra ciencia), tratando de dar una panorámica más o menor global sobre el asunto que se trata. Para ello, se suele estructurar el congreso en una larga serie de conferencias que suele durar varios días. Se destacan los conferenciantes invitados por el congreso, 10 o 20 autoridades de la materia, amigos o con ideas afines a los organizadores del congreso, a los que se les permite dar una charla larga de hasta una hora para hablar de sus investigaciones o de aquellas que le son simpáticas. Luego, están los conferenciantes seleccionados entre quienes solicitan poder impartir una charla. Dado que el número de estos conferenciantes es elevado (unos 50 en un congreso de unos 3 o 4 días sin sesiones paralelas), se reparte el tiempo que no ocupan las autoridades de modo que cada uno puede hablar unos 15 o 20 minutos. En este apretado tiempo deben contar sus investigaciones de los últimos 2 o 3 años, con lo que resulta una serie de charlas "concentradas" que agotan la atención de los oyentes en poco tiempo. Básicamente tienen una utilidad propagandística; sirven para decir: he hecho un trabajo sobre esto y el que quiera saber algo que se lea mi artículo. Finalmente está la sección de pósters, en el que un centenar de papiros condensados concentran texto y figuras en 1 metro cuadrado de cartulina cada uno para mostrar sus resultados y hacer propaganda de sus resultados. A mí, particularmente, me recuerda esto a las ferias de exposiciones en la que cada empresa expone sus mercancías con fines publicitarios. Además, cada vez más se utilizan recursos de publicista para llamar la atención de los asistentes al congreso (los mismos que exponen los pósters o dan conferencias): fotografías y diseños de pósters en colores llamativos, charlas con videos de simulaciones numéricas o películas de gran impacto visual (hay incluso casos de investigadores que pagan a creadores de animación profesionales tales como "DreamWorks" para hacer los videos), etc. Todo ello con el fin de captar la atención de los asistentes que se pierden entre las toneladas de información, de paja realmente porque tampoco hay tanto que contar de nuevo en cada congreso, simples detalles técnicos sin demasiada importancia. La batalla de los científicos no es encontrar nuevas buenas ideas, sino encontrar el modo de vender ideas mediocres. El marketing es más importante que las matemáticas. Simple publicidad, y encuentro de colegas para charlar y quizás establecer alguna futura colaboración.

   Esta publicidad es importante para el sistema, para su control del flujo de información. Y es importante dar prioridades en los congresos, partiendo ya de la base de que para asistir a ellos el científico necesita subvención que debe conseguir en su propio instituto de trabajo donde ya habrá tenido que pasar la primera purga para convencer a su proyecto de que los gastos que requiere su asistencia al congreso tienen justificación con lo que va a exponer. A veces se prohíbe la asistencia de ciertos personajes. Por ejemplo, en las conferencias de cosmología que tuvieron lugar en el Vaticano en 1982, sólo dejaron participar a los defensores acérrimos de la teoría del Big Bang, y se dejaron de lado grandes personalidades de entre los que defendían lo contrario, como Hoyle, Ambartsumian o Geoffrey Burbidge. O como relata el físico W. Kundt: "[casos] en los que no era invitado a las reuniones de los temas en que trabajo, e incluso casos donde fui enviado a casa el día de mi llegada". Afortunadamente, ni el ejemplo del Vaticano ni las experiencias que relata Kundt están demasiado extendidas. Hay un cierto filtro, sí, pero, con todo, el sistema de censura es menos eficaz que el arbitraje de las revistas y se suelen colar más teorías audaces. Claro que a la hora de realizar el libro de "proceedings" (que recoge los textos escritos de las charlas y pósters), unos llevan 2 páginas en formato pequeño o ninguna y otros llevan 30 o 40 páginas.

**Financiación, astropolíticos, supervedettes**

Dentro de los investigadores senior, no todo el mundo tiene el mismo peso o autoridad en la jerarquía. Existen, como en todo departamento universitario, ciertos rangos como el de catedrático, director del departamento, etc. Aparte de estos rangos de nombre, existe también en la práctica cierto status de poder asociado a otros factores.

Muchos de los investigadores senior, funcionarios de plaza fija, invierten la mayor parte de su tiempo dando clases en la Universidad. Si acaso, cogen algún escl... digo doctorando, para que le haga algún trabajo que él firmará conjuntamente, para que se vea que hace algo de investigación. En los institutos de mayor prestigio, donde la competencia es mayor, una parte de los investigadores son líderes de un proyecto al que aspiran a darle "impacto", es decir, que del proyecto salgan muchos artículos publicados y que estos tengan cierta importancia en el mundillo.

El investigador principal de un proyecto dirige a varios estudiantes de doctorado, varios postdoc y quizás algún que otro investigador senior de menor status. Hay incluso casos en los que este investigador principal puede tener todos los postdocs de un instituto pequeño. Suele ser una especie de gerente comercial de una empresa, con cierta parte de gestor y consejero. Yo los llamo "astropolíticos".

He conocido a muchos astropolíticos, y creo que hay un patrón de conducta semejante en la mayoría de ellos. Para hablar con ellos tiene uno que pedir cita, pues siempre están liados con mil asuntos. El "no tengo tiempo" en una de las frases favoritas del astropolítico, hombre de nuestra era. Vivimos unos tiempos en los que hasta el último mono pretende enseñorarse como si fuese un hombre importante, un ministro o qué se yo, con esa vana respuesta del "no tengo tiempo" o "estoy ocupado". Cuando uno entra en su despacho, es raro que no suenen tres o cuatro llamadas de teléfono en menos de media hora. A sus bandejas de correo electrónico llegan decenas o centenares de e-mail diarios. Cuando uno quiere fijar una fecha para mostrar unos resultados científicos y conseguir su opinión, tiene que revisar su agenda, mentalmente o en su libreta, pues siempre tiene alguna reunión en algún sitio. También muchos viajes, nacionales y al extranjero. Deben preparar diversas charlas, pues ellos son los principales ponentes de los congresos. Deben asistir a multitud de reuniones de astropolíticos para conseguir acuerdos (de colaboraciones científicas en vez de comerciales pero al caso es parecido), o negociar algunas partidas presupuestarias, o hacer propaganda del proyecto con el fin de obtener beneficios económicos u obtener impacto, o idear junto con otros algún macroproyecto que costará muchos millones de euros y que implicará el trabajo monótono de muchos investigadores (no ellos, por supuesto, sino sus súbditos y otros nuevos esclavos que comprarán con el dinero que consiguen en las negociaciones). Cuando no están de viaje ni reunidos, suelen estar ocupados elaborando los informes periódicos sobre las actividades del proyecto, o rellenando los formularios para solicitar tiempo de observación en algún telescopio (a veces esto lo hacen también los estudiantes) o solicitando ayudas económicas para el proyecto (para viajes, ordenadores, instrumental científico, etc.) al ministerio de tal o de la ayuda cual. El tiempo restante, suelen estar ocupados en la coordinación de los esfuerzos del personal del proyecto, es decir diciendo lo que se debe investigar, y escuchando (en muchos casos, no escuchan) o leyendo el trabajo de los currantes de abajo para expresar su opinión, y sugerir seguramente algún cambio. La gallina se toma un café para descansar de sus tareas burocráticas mientras los polluelos se juntan a su alrededor ansiosos de mostrar sus resultados. Rara vez suele el astropolítico sentarse a hacer el trabajo de ciencia propiamente dicho. Si acaso unas horas algún día para enseñar al de abajo, pero el estar meses y meses con un problema es cosa de sus estudiantes o postdocs.

Cuando el astropolítico tiene cualidades brillantísimas como gestor y vendedor de la ciencia que crean sus trabajadores, estamos ante un caso de "estrella" que deslumbra a sus competidores: es la "supervedette", la gran protagonista entre las protagonistas. En un instituto con más de un centenar de investigadores, sólo suele haber una o dos supervedettes. No es muy difícil identificarla porque son una referencia esencial de ese instituto, sobre todo en la imagen que da al exterior. Si viene algún periodista a hacer algún reportaje de lo que se hace en el instituto, la supervedette es primer plano. Si hace algún trabajo su grupo, rápidamente se llama a la prensa para que anuncie al mundo lo que fulanito "et al." (es decir, "y colaboradores", aunque el nombre de los currantes de abajo no suele importar y el que lleva la fama es la supervedette) acaban de descubrir. Suelen tener muy buen ojo para tocar los temas de gran impacto popular (no necesariamente temas de gran importancia científica). Y si el tema no es de bombo y platillo, ellos lo convierten en tal. Publican sin problemas en las revistas, escriben libros para científicos y para el público de la calle, son los amos de los congresos junto con otros de su igual, consiguen el tiempo de telescopio que quieran, y el presupuesto para sus actividades es bien abultado. Estas personas ya no piensan en objetivos menores sino que suelen pensar a lo grande: liderazgo en grandes proyectos multimillonarios, cobertura periodística y televisiva al alcance de una llamada de teléfono, aspiraciones a premios nacionales e internacionales como mejor investigador, codearse con grandes personalidades públicas de la alta sociedad. Dependiendo de lo gordo de la "estrella" así serán también sus circunstancias. En fin... la envidia de cualquier astropolítico.

Hay excepciones a esta conducta, claro está. Cualquier intento de generalizar el patrón de conducta de un determinado colectivo está siempre sujeto a las correcciones correspondientes por los detalles particulares de cada cual. Hay algunos casos en los que el científico senior, e incluso el líder de un proyecto, trabaja en las mismas labores que los súbditos y no dedica tanto tiempo a la gestión, sobre todo cuando logra liberarse de cargos administrativos. Sin embargo, eso no es lo más usual.

He conocido un caso peculiar de investigador senior con plaza fija que se aparta totalmente de la descripción del astropolítico. Esta persona no lidera ningún proyecto, trabaja él mismo en las ideas que tiene, o conjuntamente con alguien en alguna colaboración. Esta persona no asiste a casi ningún congreso (si acaso uno cada 5 o más años). No asiste a reuniones, ni tiene apenas llamadas de teléfono cuando uno está en su despacho, no se pasa el día contestando e-mails. Siempre que uno quiere hablar con él, tiene tiempo para recibir visitas. Aparte de trabajar con gran profesionalidad en su campo, conoce muchas otras cosas de otros campos. Esto hace que se pueda hablar con él casi de cualquier tema de física y muchos otros temas de otras áreas, pues es un buen conocedor de filosofía e historia, y tiene muy buena memoria sobre lo que lee. Está acostumbrado a pensar. De hecho, muchas veces que he llegado a su despacho lo he encontrado no haciendo otra cosa que pensar, no haciendo burocracia ni llamadas de teléfono. Cuando alguna vez he hablado con él sobre algún tema de astrofísica suelo encontrar un lúcido pensamiento en sus respuestas. Tiene un pensamiento rápido (quizás tan rápido que a veces es difícil entender lo que está pensando cuando se le escucha) y casi siempre acertado. Tiene una gran intuición para ver un problema y saber de qué va, y es fácil encontrar en él una ayuda cuando uno está atascado en algún rompecabezas físico. Esta persona es un prototipo de sabio-científico bastante raro en los tiempos de burocracia que corren. No pega con lo que se lleva actualmente; y de hecho sucede que está considerado en el status del instituto donde trabaja como un personaje de segunda fila, pocos lo conocen fuera del instituto, y sus difíciles trabajos caen casi siempre en el olvido por falta de propaganda. Hoy (y también en otros tiempos pasados probablemente) el triunfador es otro: el científico de maletín en la mano, el ejecutivo, el gestor profesional de la ciencia, que en muchos casos no sabe pensar un problema científico, o no tiene demasiados conocimientos de física y astronomía, o sus hallazgos o los de su grupo no son tan destacables desde un punto de vista objetivo (difícil es tener esa objetividad, porque para cada investigador sus trabajos son importantes), pero eso es lo de menos hoy en día para ser un científico destacado.

**Prensa, televisión, propaganda**

Como decía anteriormente, la prensa, radio, televisión o similares, son una herramienta útil para la manipulación de la información y la propaganda mercantilista. El conocimiento que la sociedad tiene de las actividades científicas procede casi totalmente de estas fuentes con lo cual constituyen un medio ideal para hacer creer lo que uno desea, siempre que uno pueda controlar esos medios.

La mayoría de los periodistas encargados de escribir los artículos sobre ciencia tienen poca idea sobre lo que escriben; si acaso poseen unos conocimientos generales de ciencias pero no pueden ni por asomo controlar todas las especialidades existentes. Esto sucede incluso en los periódicos más prestigiosos de un país. En los menos prestigiosos, hasta es probable que el periodista no tenga ni la más mínima cultura científica. Esto hace que los periodistas deban creer al investigador en lo que dice; si éste dice que acaba de hacer un descubrimiento de gran impacto, el periodista debe confiar en que así es, pues no tiene conocimientos para ponerlo en duda. Lo único que cuenta en estos casos es la reputación del investigador. De este modo la fama alimenta la fama: un investigador de prestigio suele estar rodeado de una nube de periodistas, lo que contribuirá con la propaganda que distribuyen a que su fama aumente. A tal respecto, creo que no hay mucha diferencia entre la "fama" conseguida por un científico o la conseguida por un cantante o protagonista de la prensa del corazón: todo es cuestión de empezar a salir en los medios públicos.

El investigador, aunque sobrevalorando quizás su propio trabajo, no suele deformar, o exagerar innecesariamente, o decir lo que no es, al menos no conscientemente. Suele ser el periodista quien hace todas esas cosas. El objetivo: el impacto, algo de preciado valor entre los amigos de la desinformación y el bombo y platillo (eso debe de ser lo que les enseñan en las facultades de Cc. de la información). De ahí que una buena parte de la información publicada por la prensa sobre recientes hallazgos científicos contenga importante errores y apreciaciones totalmente fuera de lugar, empezando por los titulares que deforman la noticia. Todavía recuerdo haber leído en un periódico algo como "momia extraterrestre", refiriéndose al hecho de que unos investigadores habían descubierto que el sepulcro donde estaba la tumba de algún antiguo faraón egipcio estaba construido con unas piedras de un yacimiento donde en el pasado había caído un meteorito de origen extraterrestre. Dicho sea de paso, parece ser que el tema de los extraterrestres preocupa bastante al ignorante de las ciencias y que por lo tanto los periodistas se sienten en la obligación de satisfacer a sus lectores en el ansia de noticias relacionadas con el asunto. Por ello quizás aparece tanta noticia en la prensa sobre el descubrimiento de nuevos planetas extrasolares, usualmente dando la apariencia de que es la primera vez que se encuentran tales, cuando lo cierto es que ya hay varias decenas de ellos descubiertos y todos con miles de masas terrestres, o sea que no tienen nada que ver con planetas tipo Tierra. Y, ¡no!, no se ha encontrado vida extraterrestre, que es la pregunta que siempre hacen los periodistas, hablando por boca del populacho ansioso de convertir la ciencia en un espectáculo.

Es muy grande el número de casos en el que se publica una noticia de ciencias como un gran bombazo (que la relatividad de Einstein ya no es correcta, o que hay vida en Marte, o la fusión fría, o cosas por el estilo). En muchos casos se corresponden a deformaciones que los periodistas hacen de cosas que no han entendido. En otros casos se corresponden a hallazgos reales publicados en revistas científicas, pero que todavía se están discutiendo y sobre los que hay aún cierta controversia. Pasados unos meses de la publicación sensacionalista se suele aclarar en la comunidad científica que el descubrimiento no era tal porque había algún error. Sin embargo lo que llega al público de la calle es el bombazo, no la réplica del bombazo desmintiéndolo. Por lo visto, eso no es tan comercial y no ayuda a que se vendan más periódicos o revistas.

Con todo lo imperfecta que suele ser la comunicación de la ciencia en los medios públicos (yo no aconsejaría a nadie enterarse de lo que se hace hoy en las ciencias a través de la prensa o la televisión; el que quiera saber algo de las ciencias que estudie libros de texto y que se olvide del presente; ya el futuro dirá lo que se hace ahora), ésta constituye un pilar fundamental de la relación de los científicos con la sociedad. De ella van a depender muchas de las subvenciones de sumas multimillonarias. Por ejemplo, el famoso caso que dio la vuelta al mundo de la piedra con vida de origen marciano en la Antártida que dio lugar a una fuerte subvención del gobierno americano para el tema; luego, se desmintió el asunto, la piedra estaba contaminada con vida terrestre, pero los que llenaron los bolsillos de sus proyectos ya tenían lo que querían (dicho sea de paso, el artículo publicado sobre este hallazgo por "Nature", la revista profesional de ciencias de mayor impacto, había pasado por tres o cuatro árbitros, los cuales dieron todos el visto bueno).

En otras ocasiones sucede al contrario: no son las subvenciones consecuencia de la prensa sino la prensa consecuencia de las subvenciones. Cuando se invierten cantidades ingentes de dinero en un proyecto (que pueden llegar a cientos o incluso miles de millones de euros), es necesario justificar la inversión procedente de fondos públicos. Por ello, se suele llamar a la prensa para que explique a la nación los grandes hallazgos conseguidos gracias a sus impuestos. Es nuevamente propaganda. En ocasiones hay descubrimientos relativamente importantes, pero en otras ocasiones no hay nada interesante. Especialmente en estos últimos casos se necesita a la prensa para que hinche el asunto y haga parecer que la cosa es más importante de lo que en realidad es. Se dicen cosas como que se han descubierto tantas nuevas galaxias del Universo, como si no hubiese ya millones de ellas catalogadas. El ignorante de la calle, que no sabe ni lo que es una galaxia, al leer noticias de este tipo, se convence realmente de la grandeza de la empresa acometida.

Por otra parte, la prensa no está al servicio de todos los fenómenos importantes en la ciencia. Sin fama, sin dinero y sin enchufe o el arropamiento de un grupo prestigioso de investigadores, el mejor científico que pudiera haber en un área con el trabajo más importante no sería escuchado ni se le prestaría atención. De ahí otro factor adicional para el aislamiento de los que no están dentro de la corriente del sistema.

"Lo que no se puede permitir, lo que sería escandaloso, sería que un individuo de pocos recursos pudiera conseguir lo que no consiguen otros con miles de millones de euros". Ése es el mensaje de la actual sociedad donde el capital impone su fuerza.

## Tiempo de telescopio

La astrofísica, como toda ciencia, tiene una parte teórica y una parte experimental/observacional. En el presente caso se restringe a la sola observación de la naturaleza, estamos en la ciencia de ver pero no tocar, pues no se pueden hacer experimentos con objetos astronómicos por razones obvias. Puede haber trabajos puramente teóricos que supongan un gran avance en la investigación, pero al final todo depende del contraste con las observaciones. Toda teoría, para tener éxito frente a sus competidoras, debe poder predecir ciertos fenómenos que otras teorías competidoras no pueden. Hasta la misma teoría de la relatividad general de Einstein tuvo que esperar a la confirmación observacional de la curvatura de la luz de las estrellas por el campo gravitatorio solar, medida en un eclipse de Sol en 1919 por Eddington et al. (realmente, fue uno de los súbditos, Crommelin, junto con otros colaboradores, quien hizo las medidas con mayor precisión desde Brasil, mientras que el grupo de Eddington en la Guinea española tuvo peor tiempo y no consiguió unas medidas tan buenas) para que tuviese la repercusión que ha tenido. Por ello, el avance de la astrofísica está íntimamente relacionado con el avance de las observaciones.

La astrofísica, y creo que la ciencia en general, tuvo a principios del siglo XX importantes auges debidos en gran medida a la búsqueda de varios investigadores de nuevas ideas. Cuando se le preguntó a Hubble y Eddington qué esperaban encontrar con los nuevos telescopios de 5 metros que se iban a construir, ellos respondieron: "si conociésemos la respuesta, no tendría propósito el construirlos". Hoy, sin embargo, la situación es bien diferente. Los grandes telescopios, y aun los no tan grandes, son utilizados sólo en el caso de que uno convenza a un tribunal de expertos de que el uso que se le va a dar es con fin de hallar algo que se espera encontrar. Para utilizar las grandes instalaciones, es preciso rellenar algunos formularios entre 6 y 12 meses antes de la fecha de observación. En esos formularios se debe dejar claro lo que uno espera encontrar con las observaciones solicitadas. Se debe mostrar al tribunal de expertos cuál es el propósito de las observaciones y qué se pretende demostrar con ellas. Aparte, deben figurar datos referentes al perfil del investigador o investigadores. Ni que decir tiene que cuanto más publicaciones observacionales tenga un investigador y más tiempo de telescopio haya conseguido en el pasado, más probabilidades tendrá de que le vuelvan a dar más tiempo de telescopio, con lo cual este investigador conseguirá publicar más artículos que nadie, aunque sean todos parecidos y sin ninguna idea de valor, y su prestigio irá en aumento. El hecho mismo de haber utilizado un gran telescopio o satélite para obtener unos datos da también prestigio a los resultados publicados. Se dice, por ejemplo, "datos obtenidos con el telescopio espacial Hubble", y con ello se presume de que el valor de que lo que estos datos aportan a la ciencia es mayor que cualquier otra información recogida con telescopios de menos fama. Es una pescadilla que se muerde la cola; el caso es entrar en el círculo, y para ello es preciso ganarse ciertas simpatías con los ya establecidos. De mi propia experiencia enviando propuestas y otras que me constan, he encontrado también que la probabilidad de conseguir tiempo de telescopio aumenta muy significativamente cuando en la propuesta figura uno de los miembros del tribunal (aunque ese miembro mismo no vaya a juzgar esa propuesta en concreto).

Con todo, el problema más grande para el avance de la ciencia está en que las nuevas ideas suelen estar mal vistas entre los tribunales que conceden tiempo de telescopio. Si uno solicita el uso de un telescopio con el fin de comprobar cómo se ajustan las predicciones de una teoría alternativa a la comúnmente aceptada, lo más probable es que se deniegue la observación. No estamos hablando de aficionados a los que circunstancialmente se les pase una majadería por la cabeza; estamos hablando de grandes profesionales, cuyo único defecto es dudar de las ideas que todos los demás creen intocables. El sistema no está por la labor de sostener una pluralidad ideológica dentro de una ciencia. Se habla de una libertad en la investigación, pero ello no deja de ser una mentira como tantas otras que pregonan los políticos en nombre de la democracia: claro que cada uno puede pensar lo que quiera, pero las instalaciones, las publicaciones de prestigio y la propaganda son sólo para los que quieren hacer de la ciencia una reconfirmación de ciertos prejuicios antes que la apertura a nuevos horizontes.

Puede ocurrir también que alguien presente una idea o un objeto interesante para su estudio, pero no puede desarrollar su trabajo porque el tribunal no le concede tiempo de telescopio. Enseguida, la gente del tribunal que tiene medios y ve que la idea tiene interés, se pone a desarrollarla al margen de las personas que la tuvieron en un principio, haciendo el descubrimiento suyo.

Todo esto es comprensible, aunque no aceptable, al menos bajo mi punto de vista. Esta política científica es conveniente cuando se trata de movimientos de grandes capitales. Un telescopio como el que ahora se está construyendo de 10 metros en La Palma con presupuesto español cuesta la tremenda cantidad de unos 100 millones de euros. Claro que puestos a buscar cifras altas, los telescopios espaciales, y los grandes proyectos en satélites para la investigación van más allá de los 1.000 millones de euros, subvencionados por varios países conjuntamente. Aparte de la construcción están los gastos de mantenimiento. En conjunto, echando cuentas de la duración media de un gran telescopio terrestre o espacial, cada hora de observación en uno de estos cacharros cuesta miles o decenas de miles de euros. Parece lógico por tanto no darle tiempo de observación al primer chiflado que aparezca que quiera vender el oro y el moro. Además, a la hora de conseguir estas inmensas fortunas que maneja la ciencia de hoy en día, es mejor dar una imagen de una ciencia firme que sabe a donde va y que tiene claro cuáles son los problemas que están resueltos y cuáles los que quedan por resolver a base de poner dinero para los aparatos. No interesa dar una imagen de una ciencia plural, dividida, inmersa en discusiones sobre ciertos fundamentos. No se invierten estas millonadas para que los científicos jueguen a adivinar cómo es la naturaleza; se invierte para obtener un producto firme lejano de la verborrea especulativa de los filósofos. En otras palabras, se está comprando la ciencia, y quien paga tiene derecho a exigir que los frutos sean superiores a los productos de menor precio. Todo se contabiliza al respecto: número de publicaciones obtenidas con un telescopio, número de citas obtenidas por esos artículos (lo que llaman el "impacto", que es un parámetro tan relacionado con la calidad de un trabajo como lo son las cifras de espectadores en los diversos programas de las cadenas de televisión), etc. Al final los informes deben hablar de cuán rentable ha sido la inversión. Parámetros como la genialidad, la creatividad, la lucidez mental y otros factores humanos escapan a esos informes.

No cabe hablar de espontaneidad en la ciencia actual, o de descubrimiento fortuito auspiciado por sospechas de que algo no va por buen camino en la astrofísica. Casi todo está planeado para que transcurra dentro de las programaciones y previsiones a años vista (desde que se comienzan los planes de construcción de un satélite hasta que se lanza pasan unos 15 años). La previsibilidad es poco común a lo largo de la historia de la ciencia, donde numerosas veces hubo que desandar algún camino para tomar nuevas vías o enfoques, o donde azarosamente se encontraba uno con resultados sorpresivos. Sin embargo, el sistema contemporáneo parece estar más seguro que nunca de no caer en errores históricos; al menos, si hubiese algún error, se trata de retrasar su descubrimiento lo más posible. Parece paradójico, pero a veces las grandes posibilidades de la ciencia para observar y experimentar pueden obstaculizar más que motivar el avance de la misma. ¿Va la astronomía para atrás con el avance de la tecnología? En cierto modo no, pues está fuera de toda duda que la astronomía es una ciencia observacional con mayores "posibilidades" cuanto mejores son los instrumentos, pero creo que, en cuanto a que el control del sistema sobre la ciencia es mayor cuanto mayores son los telescopios, bloqueando así iniciativas privadas fuera de lo comúnmente aceptado, sí va para atrás. Es decir, las "posibilidades" crecen pero el uso eficiente de esas posibilidades decrece. Además, los grandes telescopios no piensan por sí solos, y el avance de la tecnología parece producir atrofia mental en algunos sectores de la ciencia.

## El avance/estancamiento de la ciencia

Se pueden dar muchos ejemplos reales, muchos nombres, y sacar a la luz muchos problemas de la astrofísica que han sido manipulados a favor de una determinada tendencia, eliminando prejuiciosamente alternativas opuestas. No es cuestión aquí de detenerse a discutir teorías concretas. Puedo hablar, no obstante, de algunos casos de tendencias generales. Por ejemplo: el uso casi único de la gravitación para entender la mayor parte de los problemas astrofísicos a gran escala: cosmología, dinámica galáctica, formación de estructuras a gran escala en el Universo, etc. Hay una corriente bastante fuerte en esta dirección. Hay alternativas, claro que las hay, por ejemplo el caso de las interacciones electromagnéticas a gran escala, pero tratar con estas fuerzas supone un esfuerzo mucho mayor que tratar con las fuerzas gravitatorias. Las incertidumbres que tenemos sobre los campos magnéticos intergalácticos, por ejemplo, son enormes. ¿Qué se hace entonces? Se pasa del problema. Se trata de resolver cualquier problema astrofísico a base de gravedad y cuando se menciona el tema de los campos magnéticos aparecen caras raras y expresiones como diciendo "no nos compliques la vida, que nosotros estamos contentos con lo que hacemos". Sin embargo, la naturaleza puede ser complicada de conocer, y el hecho de que algunos no quieran complicarse la vida no tiene por qué llevarnos a un conocimiento verdadero. Hay quien intenta hacerse el filósofo citando aquello de la navaja de Occam (alguno parece que es la única cita filosófica que conoce), pero mal citado, porque se confunde la simplicidad de la naturaleza con la simplicidad de lo que ellos pueden calcular. "A la naturaleza no le importan las dificultades analíticas"—decía Fresnel en 1826. Hay muchos problemas de la astrofísica que no se resuelven limpiamente con las interacciones gravitatorias; y ante eso, los líderes de la investigación llevan a la ciencia por caminos especulativos con términos tales como: agujero negro supermasivo, materia oscura no bariónica, inflación, constante cosmológica, lente gravitatoria...; algo que ni ellos mismo comprenden a veces, pero que les permite seguir hablando en términos gravitatorios y evitar tener que aprender nuevas ramas de la física que no sean aquellas a las que han dedicado 20 o 30 años de su vida. A grandes problemas, como el de la luminosidad de los cuásares, grandes agujeros negros que nadie ve, de millones o decenas de millones de masas solares. Aquí es donde se olvidan de su navaja de Occam y no les importa poner los parches que se necesiten a sus prejuicios con tal de no abandonarlos. Lo mismo sucede con respecto al tema de la extinción intergaláctica (de la que ante falta de conocimientos exactos se toma a conveniencia como nula en todas las longitudes de onda hasta muy largas distancias), o la distancia de los cuásares (sobre la que hay un común acuerdo de tomar en todos los casos la distancia cosmológica derivada del desplazamiento espectral al rojo, a pesar de los problemas que ello conlleva para explicar algunas correlaciones encontradas entre galaxias cercanas y cuásares distantes) u otros problemas ante los cuales muchos miran hacia otro lado cuando se los mencionan. ¿Falta materia para justificar una dinámica que no concuerda con la materia que vemos? No hay problema, decimos que hay materia oscura y ya está. Avanzan las investigaciones y se ve que esa materia no puede ser ningún tipo de materia conocido. Entonces, se vuelve a poner a otro parche a la teoría y se inventan nuevos tipos de materia nunca vistos: materia no bariónica, que también sirve para resolver el problema de encontrarse que las anisotropías de la radiación cósmica de fondo son mil veces menores de lo que se esperaba antes de medirlas. Y se inventa la inflación, y se quita y se pone la constante cosmológica de las ecuaciones de Einstein a antojo, según la moda, y se añaden más y más parámetros libres a una teoría de manera que si algo no encaja con los datos, es cuestión de retocar unos cuantos parámetros ad hoc. Etc. ¿Y hasta cuando esta tendencia de agarrarse a una teoría, ponerle parche sobre parche, e ir enderezándola para que prediga a posteriori cualquier cosa? Pues hasta que alguien se dé cuenta de que hay algún fallo de base en el edificio hecho a pedazos, y entonces es tiempo de tirarlo todo y empezar de nuevo en alguna otra parte. Ejemplo histórico bien claro: la revolución copernicana que tiró abajo la archiparcheada astronomía de Aristóteles-Ptolomeo. Esto es precisamente lo que se trata de evitar a toda costa.

Quizás todas las teorías alternativas están equivocadas, quizás. Aún así, ¿podemos estar 100% seguros de que los escenarios estándar son correctos como para rechazar sistemáticamente todas las propuestas alternativas solamente porque van en contra de la visión ortodoxa? Creo que no. Sin embargo, el sistema actúa aparentemente como si tuviera la teoría final en sus manos. El sistema posee una colección de modernas teorías parcheadas, tal como la de Aristóteles-Ptolomeo, y teme que se pierda su status. Galileo no lo tuvo fácil en su tiempo para luchar a contracorriente, y hoy la historia sigue siendo la misma. Parece que nada cambia. Lo penoso del asunto es que hoy se nos quiera vender una imagen de libertad alejada de las circunstancias de hace cuatro o cinco siglos, y lo cierto es que sigue habiendo los mismos perros con distintos collares. Al menos no llevan a nadie a la hoguera, es cierto, a lo sumo te exilian del reino que les pertenece: te hacen la vida imposible para que no publiques o avances en tus investigaciones; algo hemos progresado. Lo penoso del asunto es también que luego se venda una imagen de la cosmología, por ejemplo, como un asunto en el que ya se tiene claro casi todo en cuanto a conceptos y sólo falta afinar algunos parámetros. Probablemente, toda la base del edificio de la cosmología actual, ciencia relativamente reciente si es que se le puede llamar ciencia, esté totalmente equivocada. Sin embargo se sigue construyendo con rapidez sobre arenas movedizas. Parece que no aprendemos nada de la historia. Parece que los intereses económicos que mueve el negocio de la ciencia precisan de una imagen de conocimiento robusto y que camina firmemente por el buen camino y eso se antepone a cualquier sabotaje por parte de algún elemento crítico disconforme.

Cuando se dialoga sobre esto con un ortodoxo, la respuesta suele ser que la ciencia oficial es objetiva, y si se ha apoyado más una teoría que otra es porque se han conseguido pruebas que apoyan a la primera antes que a la segunda. Esto, dicho así, suena muy bonito y hasta parece honesto. Sin embargo, a la luz de lo que he dicho a lo largo del presente artículo, hay que considerar que no todo es tan limpio. No digo que sea todo pura manipulación, no, hay muchas ocasiones en que la naturaleza se muestra clara en los experimentos u observaciones y las conclusiones son efectivamente claras. Sin embargo hay muchos casos turbios, pertenecientes a ciencias turbias (como la cosmología) en que el poder de manipulación puede más que la misma realidad.

Es cierto que ciertas teorías estándar funcionan mejor a la hora de explicar los datos. Lo que no se cuenta muchas veces es la cantidad de personas que trabajan en esa teoría, poniendo parches aquí y allá. Se dice que la teoría del Big Bang, por ejemplo, ha vencido a sus competidoras. Claro, esta teoría cuenta con varios miles o decenas de miles de investigadores que de uno u otro modo están implicados e interesados en que la teoría sea correcta para que el trabajo de toda su vida no se vaya a pique. Una teoría de la oposición puede ser defendida por un número de investigadores que se cuentan con los dedos de las manos; por muy buenos que éstos sean, no pueden competir en producción de parches ni en campaña publicitaria con las ingentes masas de investigadores de lo ortodoxo. Estos investigadores tienen incluso que luchar contra el sistema sin dinero, sin estudiantes, sin tiempo de telescopio. Personalmente, pienso que la cosmología no es una ciencia seria, y no me creo ni el Big Bang ni las teorías de la competencia. En cualquier caso, desde un punto de vista deportivo, me parece que, en estas condiciones, el juego no es muy justo y no parece apropiado hablar de derrota sino más bien de abuso.

Se suele apelar a la objetividad del método científico. Sin embargo, en asuntos turbios, el método que se suele aplicar no es demasiado objetivo. Básicamente es el siguiente:

- Sea una teoría A autodenominada ortodoxa o estándar, y una teoría B no-ortodoxa o no-estándar. Si las observaciones apuntan a lo que predecía la teoría A y que no predice B, supone un gran éxito de la teoría A, algo que debe difundirse inmediatamente en todos los medios de comunicación importantes. Supone que ya no cabe duda de que la teoría A es la correcta. La teoría B es incorrecta; uno debe olvidarse de esa teoría, y por tanto bloquear cualquier investigación en ese aspecto (poner obstáculos para la publicación, que no se conceda tiempo de telescopio, etc.).

- Si las observaciones apuntan a lo que predecía la teoría B y que no predice A, esto no supone nada. La ciencia es muy compleja y antes de adoptar una posición debemos seguir pensando en el asunto y hacer más comprobaciones. Probablemente el observador se ha equivocado, se necesitan más observaciones (que difícil será que las hagan porque no les vamos a dejar los telescopios para que comprueben su estúpida teoría B). ¡Quién sabe! Quizás lo que se observa es debido al efecto "Pepito", claro; quizás no han corregido este efecto del que no sabemos nada. Todo es tan complicado. Debemos estar seguros antes de poder decir algo acerca de qué teoría es la correcta. Además seguro que A también puede predecir lo mismo, añadiendo algunas reformas a la teoría (como tenemos un ejército de teóricos dispuestos a poner parches y descubrir nuevos efectos, en menos de tres meses seguro que podemos crear una teoría A con ciertas reformas que sea consistente). En cualquier caso, ante el río revuelvo, mientras no nos aclaramos, nos quedamos con la teoría A. Quizás, como decía Halton Arp, el dicho informal "para hacer cambios extraordinarios se requieren evidencias extraordinarias" realmente quiera decir "para hacer cambios que personalmente no me convienen ninguna evidencia es suficientemente extraordinaria".

**Ciencia extraoficial**

El sistema invita realmente a ser abandonado y realmente estoy convencido de que si alguien quiere hacer algo importante (aquí nuevamente insisto en que no sólo es el caso de las ciencias sino, generalmente de cualquier mafia humana con nombre de "cultura") debe hacerlo al margen de la oficialidad, quizás en el tiempo libre, quizás con el estudio afanoso por cuenta propia. El problema de esta posición en cuanto a las ciencias está en las precarias o nulas posibilidades de que se dispone para la observación y la experimentación, y en lo desprestigiado que está el librepensamiento al margen de las instituciones oficiales. Dados los altísimos costes del material necesario para la investigación empírica puntera en cada campo, queda anulada la posibilidad de ejercer tal profesión desde la periferia. Sólo queda la posibilidad de la teorización/especulación pura, o quizás alimentada de los datos empíricos producidos por los demás científicos. Eso, de hecho, se da con mucha frecuencia.

Con bastante frecuencia (casi cada mes o cada dos meses) suelo recibir por correo electrónico nuevas teorías de aficionados que tratan de derrumbar todo lo bien conocido para dar lugar a nuevas teorías físicas de carácter ridículo o teorías cosmológicas que fallan hasta en lo más básico. Este tipo de trabajos carece prácticamente de referencias especializadas, citando más bien obras de divulgación, y en vez de centrarse en un problema concreto de la ciencia, tratan de asuntos muy generales, tratan de derrumbar de un plumazo toda la física conocida, por ejemplo. Ése es precisamente el problema de la credibilidad de la investigación con independencia de las instituciones oficiales: por cada individuo que quiere tomarse la profesión en serio y con una preparación adecuada, existen miles de chalados por el planeta que sueñan con crear una teoría física inspirada por el cielo como quien escribe poesía, y que arrase con todo lo pasado e instaure una nueva era en la historia de las ciencias. Los objetos exóticos, como los agujeros negros, son también de gran interés de los aficionados. Una vez oí en la radio a un carpintero de profesión que nunca había estudiado física y que tan sólo había leído algunos libros de divulgación, sin tratar de entender nada de las matemáticas del asunto; el hombre decía que había escrito 7 libros sobre agujeros negros, y se quejaba de que no le habían publicado ninguno. No quiero juzgar negativamente esos esfuerzos de los "aficionados", que quizás se han leído algún libro de divulgación de Hawking y ya creen que están en disposición de trabajar como investigadores. No quiero actuar yo como esa parte del sistema que castra cualquier intento de originalidad por el mero hecho de ser desafiante. Sin embargo, la realidad es que todas esas teorías que a los aficionados se les ocurren están llenas de fallos e inconsistencias, por falta de conocimientos, independientemente de su ideología. Y el problema es precisamente que los miles de chiflados que hay en la penumbra no dejan escuchar la voz de algún posible genio que pueda haber en el medio. Es por ello que las actividades desde el punto de vista autónomo no tienen una buena prensa y uno tiene que recurrir a los mecanismos oficiales si quiere ser escuchado por alguien del gremio.

**La actitud de los filósofos ante la ciencia**

Es probable que alguien identifique el presente manifiesto como una crítica filosófica, de las tantas que se suelen cargar contra la ciencia. Creo que hacer una reflexión del tipo de lo que hago sí es hacer filosofía, sin embargo no ha de confundirse el presente texto con el tipo de producciones por los autoproclamados filósofos profesionales (los que tienen título, para entendernos).

Lo cierto es que no es frecuente encontrar muchas críticas de este tipo al sistema de la ciencia. Hay críticas, sí, pero alejadas de un contacto real con el problema y parafraseando especulaciones metafísicas que bien poco tienen que ver con los problemas a que me refiero. La filosofía de la ciencia de hoy se enseña en las facultades de filosofía es realmente una filosofía de la anticiencia, se enseña que los científicos son unos ineptos que no saben pensar y que son ellos, los filósofos profesionales, los que son llamados a dar un sentido a las ciencias y su significado. Hay varios líneas dentro de la filosofía de la ciencia, y no voy a hacer aquí un recorrido extenso sobre las mismas pues no es la misión de este artículo. Hay un amplio abanico: desde las posiciones más abiertamente anticientíficas (que comparan a la ciencia con una religión y piensan que no existe ninguna verdad en nuestros conocimientos científicos, que tanto valor tiene la ciencia de un brujo de una tribu africana como nuestra ciencia occidental); hasta las más tranquilas que se limitan a escribir libros gordísimos explicando al mundo qué es una hipótesis, en qué consiste el método científico, la falsabilidad de una teoría, y esas trivialidades que cualquier científico conoce bien desde su temprana formación y que a nadie van a desvelar nada que no se sepa. Fuera de esto, existe muy poco esfuerzo por parte de los filósofos profesionales por entender los problemas que tiene la ciencia actual. Si acaso se cita puntualmente alguno de estos problemas con el fin de desprestigiar la ciencia de todos los tiempos en su conjunto, para luego terminar ellos diciendo: "claro, ya lo decíamos nosotros...", y seguidamente tratar de vender alguno de sus credos metafísicos para-anormales que son mercancia usual entre los sofistas modernos. Ya se sabe lo que dice el refrán: "del cielo para abajo, cada uno vive de su trabajo".

¿Por qué no es frecuente este tipo de críticas entre los trabajos de los filósofos profesionales? Pienso que hay dos razones fundamentales: 1) que no conocen la ciencia de cerca sino a través de la lectura de libros donde no vienen reflejados muchos problemas reales; 2) que no les interesa descubrir problemas en otros gremios dado que ellos mismos tienen esos problemas, si acaso ampliamente magnificados. Los problemas de la obstaculización de la libre iniciativa en una línea de investigación o ideología están presentes en las facultades de filosofía más que en las de ciencias. Los congresos de filósofos no son sino meras imitaciones de los congresos científicos. La censura en las publicaciones, la mayoría de carácter pueblerino en vez de internacional, es mucho más evidente; ante la falta casi total de criterios objetivos y de datos empíricos, se publica o se rechaza un artículo a antojo sin ni siquiera molestarse a emitir un informe para las causas del rechazo. Las plazas se dan casi a dedo. La comunicación con la prensa, o la publicación de libros por parte de las editoriales está casi exclusivamente al servicio de las supervedettes correspondientes. La propaganda es la que decide la supervivencia de una majadería filosófica. Etc. Con este panorama, ¿qué tienen que decir los filósofos de despacho sobre la ciencia? No es de extrañar pues el silencio en estos aspectos y que concentren sus esfuerzos en escudriñar qué es la verdad o cuántos tipos de razón existen o los muchos ejercicios de análisis del lenguaje y la clasificación de las escuelas en los diferentes "-ismos". Como decía, del cielo para abajo cada uno vive de su trabajo, y los problemas de la ciencia no los va a encontrar ni solucionar un filósofo a sueldo. Es cosa de los que viven la ciencia desde dentro el cuestionárselo.

## Unas notas de optimismo finales

En definitiva, veo con cierto pesimismo el estado actual de la astrofísica, así como el de otras ciencias, que tendrán problemas semejantes. Todas las circunstancias descritas pueden llevar a pensar, yo así lo veo, que la ciencia actual en un ramo como el de la astrofísica está bastante prostituida. Esto es lo que la astrofísica y el oficio más viejo del mundo tienen en común. Estos problemas se reflejan en muchas otras áreas de la cultura, y en la sociedad misma. Todo se contamina, no son problemas aislados: cada actividad humana es un reflejo de lo que le circunda. Vivimos en una sociedad podrida que se engaña a sí misma, ¿qué podríamos esperar de la ciencia que de ella sale?

Debo reconocer que mi crítica tampoco es objetiva. Quizás mi pesimismo sea exagerado, en contraposición al exagerado optimismo de otros. Quizás mi visión es algo desproporcionada. En fin, cada cual que juzgue por sí mismo; yo sólo expreso una opinión y cada cual tendrá la suya. Esto no es un panfleto con fines políticos o sectarios, no estoy interesado en convencer de nada. No creo que el presente texto sea útil para fines sindicalistas que reclamen los derechos de los trabajadores de la ciencia. Aquí no se trata de derechos, sino de hechos: de conocer la naturaleza como es. De lo que se trata es de hacer una ciencia de calidad. De nada sirve apelar a unos "derechos" porque seguro que los problemas de la investigación científica no se solucionarán en nada con el incremento de burocracia sino todo lo contrario. Tampoco se trata de señalar los problemas existentes ante falta de presupuesto dedicado a la ciencia. Todo lo contrario, cuanto más dinero hay por medio más mafioso se vuelve el sistema. Pesimista me muestro por tanto hasta para buscar soluciones políticamente correctas en el contexto social actual.

No obstante, no quiero dejar de comentar en estas líneas finales que no todo es de color negro y que hay ciertos motivos para estar, sino plenamente satisfecho, al menos con cierto grado de optimismo. A decir verdad, a pesar de la cantidad de problemas que existen en la infraestructura de la ciencia, creo que se puede hacer ciencia y que hay un cierto avance neto; no un avance descomunal tal como nos da a entender la propaganda, pero sí un cierto avance; lento, con muchos errores que se corrigen muy lentamente, pero hay cierto avance y nuestros conocimientos sobre el Universo crecen. Otras épocas también tuvieron múltiples problemas para lograr vencer las dificultades, pero al final parece que hay ciertos mecanismos de la historia por los cuales la ciencia se va puliendo y se van filtrando conocimientos más sólidos con independencia de intereses humanos. Probablemente, porque los intereses creados se van disolviendo con el paso de las generaciones, y al cabo de unas decenas de años o quizás unos siglos, sólo pueden sobrevivir las ideas que tienen algún valor por sí mismas. Realmente, no siempre la historia es justa. Muchas buenas ideas pasan al olvido y no se recuperan hasta que vuelven a ser pensadas de modo independiente. Copérnico tuvo que redescubrir lo que Aristarco de Samos sabía ya 17 siglos antes. Son múltiples los casos de investigadores que la historia hace famosos, perteneciendo el mérito a unos desconocidos. Tampoco la historia es perfecta, es humana al fin y al cabo. Con todo, creo realmente que hay algo grande en la astronomía, en la física, en todas las ciencias de la naturaleza que permiten al hombre salir de su agujero y entender parte de aquello que supera todas las insignificantes mezquindades del animal humano. Hay una naturaleza, hay un Cosmos, y nosotros caminamos hacia su conocimiento, ¿no es eso algo maravilloso? Los encantos de la profesión son muchos, como encantos tiene el amor siempre y cuando no se pongan al servicio de fines mercantiles.